\documentclass{aa}  

\usepackage{graphicx}
\usepackage{txfonts}
\usepackage{verbatim}
\usepackage[utf8]{inputenc}

\begin{document} 

\title{Deep search for hydrogen peroxide toward pre- and protostellar objects}

\subtitle{Testing  the pathway of grain surface water formation}
\titlerunning{Search for HOOH toward YSOs} 
\author{G.W. Fuchs
          \inst{1}, 
                 D. Witsch \inst{1},
                 D. Herberth \inst{1},
                 M. Kempkes \inst{1},
                 B. Stanclik \inst{1},
                 J. Chantzos \inst{2},
                H. Linnartz  \inst{3},
               K. Menten  \inst{4},
        \and 
              T.F. Giesen \inst{1}
%\fnmsep\thanks{Just to show the usage of the elements in the author field}
          }
\authorrunning{G.W. Fuchs et al. }

\institute{Institute of Physics, University Kassel,
              Heinrich-Plett Str. 40, 34132 Kassel, Germany\\
              \email{fuchs@physik.uni-kassel.de}
         \and
             Max-Planck-Institute for Extraterrestrial Physics (MPE), Giessenbachstraße 1, 85748 Garching,  Germany\\
        \and
            Laboratory for Astrophysics, Leiden Observatory, Leiden University, PO Box 9513, NL2300 RA Leiden, The Netherlands.\\
        \and
             Max-Planck-Institut for Radio Astronomy (MPIfR), Auf dem Hügel 69, D-53121 Bonn, Germany.\\
             %\thanks{The university of heaven temporarily does not  accept e-mails}
             }

\date{Received December 14, 2018; ...}

% \abstract{}{}{}{}{} 
% 5 {} token are mandatory

\abstract
  % Context heading (optional)
   {In the laboratory, hydrogen peroxide (HOOH) was proven to be an intermediate product in the solid-state reaction scheme that leads to the formation of water on icy dust grains.
When HOOH desorbs from the icy grains, it can be detected in the gas phase. 
In combination with water detections, it may provide additional information on the water reaction network. 
 Hydrogen peroxide has previously been found toward $\rho$ Oph A. 
However, further searches for this molecule in other sources failed. 
Hydrogen peroxide plays a fundamental role in the understanding of solid-state water formation and the overall water reservoir in young stellar objects (YSOs). 
Without further HOOH detections, it is difficult to assess and develop suitable chemical models that properly take into account the formation of water on icy surfaces.
}
   % Aims heading (mandatory)
   {The objective of this work is to identify HOOH in YSOs and thereby constrain the grain surface water formation hypothesis. 
    } 
   {
Using an astrochemical model based on previous work in combination with a physical model of YSOs,  the sources R CrA-IRS\,5A, NGCC1333-IRAS\,2A, L1551-IRS\,5, and L1544 were identified as suitable candidates for an HOOH detection. 
Long integration times on the APEX 12m and IRAM 30m telescopes were applied to search for HOOH signatures in these sources.
}
   % results heading 
{None of the four sources under investigation showed convincing spectral signatures of HOOH. 
The upper limit for HOOH abundance based on the noise level at the frequency positions of this molecule for the source R CrA-IRS\,5A was close to the predicted value. 
For NGC1333-IRAS 2A, L1544, and L1551-IRS\,5, the model overestimated the hydrogen peroxide abundances. 
}
 {HOOH remains an elusive molecule. 
With only one secure cosmic HOOH source detected so far, namely $\rho$ Oph A, the chemical model parameters for this molecule cannot be sufficiently well determined or confirmed in existing models. 
Possible reasons for the nondetections of HOOH are discussed. 
    }

\keywords{hydrogen peroxide (HOOH) --
                     water formation --
                     grain surface/ gasphase reactions -- 
                Young Stellar Objects (YSO) --
                pre- and protostellar objects --
                chemical-physical model -- 
               L1544 --  L1551-IRS\,5 --  NGC1333-IRAS\,2A -- R CrA-IRS\,5A 
               }

\maketitle
%
%________________________________________________________________

\section{Introduction}

Water is the essence of life and is one of the most relevant molecules in the Universe. 
In star formation processes, H$_2$O plays an important role as coolant; in dense cold regions, water is the dominant constituent in ices coating cosmic grains; and generally, water can be used as a valuable physical diagnostic and evolutionary tracer \citep{vanDishoeck2011}.  
Its origin and abundance in space is at the core of fundamental questions that still lack conclusive answers. 
Three reaction schemes are commonly used  to explain water formation in space \citep{vanDishoeck2013, Hollenbach2009}. 
In hot or radiation-exposed environments, gas-phase reactions dominate that are either based on high temperature (1) or ion-molecule reactions (2). 
In cold and radiatively well-shielded regions where ices form on cold dust grains, solid-state reactions (3) are at play. 
The latter mechanism is believed to dominate in the environments of star-forming regions. 
Dedicated observing programs such as the Herschel-WISH program \citep{vanDishoeck2011} have investigated the abundance and occurrence of water in our Universe and in our local galactic environment. 
In addition, laboratory data of chemical processes in interstellar ice analogs became available that helped to better understand the mechanisms of water formation on interstellar dust surfaces \citep{Ioppolo2010}. 
% {GWF: I now strat  the sentence without a ``The `` 
Laboratory studies showed that hydrogen peroxide (HOOH) acts as a prominent intermediate in the hydrogenation chain from oxygen allotropes (O, O$_2$ , and O$_3$) to the formation of water \citep{Ioppolo2008, Miyauchi2008, Oba2009, Dulieu2010, Cuppen2010, Romanzin2011}.
This also means that HOOH has the potential of acting as a tracer of solid-state reactions that result in the formation of water. 
%{GWF: I have made two sentences out of one. See next two sentences.}
This depends on the amount of HOOH that is present in the ice because it may fully react to form water, and it depends on the efficiency with which HOOH can be released into the gas phase after it is formed, for instance, through reactive desorption or dissociative desorption processes. 
In addition, the amount of observable HOOH may be reduced by  gas phase processes that consume HOOH once it is released from the solid. 
% {GWF: I replaced $H_2O_2$ with HOOH and added something to be more specific 
The first point, that is the amount of solid-state HOOH,  has been addressed by \citet{Smith2011} by investigating interstellar ices at 3.5, 7.0, and 11.3$~\mu$m where HOOH can be distinguished from water ices; see also \citet{Boudin1998}. 
% {GWF: I replaced $H_2O_2$ with HOOH
However, no conclusive evidence for HOOH as grain mantle constituent could be reported.  
At best, the 3.47$~\mu$m feature at the shoulder of the long-wavelength wing  of the 3.08$~\mu$m water O-H stretching feature might be assumed to originate from HOOH (but possibly also from other molecules) and might thus be used to infer an upper limit, which would then be between 2\% and 17\%. 
However,  already in 2011, gas-phase HOOH was identified toward the star-forming region $\rho$ Oph A \citep{Bergman2011a}.  
\citet{Du2012} developed a detailed model based on chemical surface reactions and subsequent desorption processes to explain these observations.  
%{GWF: I made change in the folowin sentence. I don't know what the exact time line was, thus I simply use ``subsequently'' }
They also predicted that HO$_2$  exists in $\rho$ Oph A in sufficient amounts that were subsequently detected by \citet{Parise2012}.  
Based on these first successes, a survey toward various inter- and circumstellar environments, including star-forming regions at various phases of stellar evolution, was started by Parise and colleagues to search for further sources of hydrogen peroxide. 
Surprisingly, it was not possible to identify HOOH in any of these targets \citep{Parise2014}.  
\citet{Liseau2015} reported a tentative detection of hydrogen peroxide toward OMC-1, but because the millimeter spectrum in this source is very dense, a firm assignment is still pending. 
Because water can be seen in many star-forming regions, including those investigated by Parise {\em et al.,} it remains puzzling why HOOH is so difficult to observe.
Parise {\em et al.} argued that $\rho$ Oph A appears to be a somewhat special source with favorable conditions for an HOOH detection mainly because the bulk of gas can be found under constant conditions at moderate temperatures of around 20-30 K.  
The heating source responsible for these temperatures is the nearby star S1 (a close binary system of the classes B4 and K).
It has also been pointed out that sources in which the molecule O$_2$ has been observed would be ideal candidates for HOOH detections \citep{Liseau2015}
because molecular oxygen is a direct precursor molecule for HOOH on grain surfaces \citep{Cuppen2010}. 
However, O$_2$ is difficult to detect, and it is commonly believed that it is not a major repository of oxygen in molecular clouds. 
%{GWF: With the changes I found the sentence less readable. What I intended to say is that O2 (oxygen) has not been found either as gas-phase nor as solid state molecule}
In only two sources has molecular oxygen been identified so far, namely  $\rho$ Oph A and Orion A \citep{Larsson2007, Liseau2012, Goldsmith2011, Chen2014}, and many further attempts to find it in the gas phase have failed \citep{Goldsmith2000, Melnick2005, Yildiz2013, Wirstrom2016}, as did the attempts to find it in solid form \citep{VandenBussche1999, Tielens2000}.

%{GWF: I tried to follow your suggestion concerning the citation}
In the model by \citet{Du2012} a desorption mechanism is included that is based on the release of excess energy during the molecule formation. This concept has also been applied by
\citet{Cazaux2016} and \citet{Minissale2016}. 
Initially, and already including gas- and solid-phase reactions, \citet{Du2012} constrained themselves to stationary and constant physical conditions in their model. 
Later, they applied the model to account for spherical symmetric problems such as in NGC 1333-IRAS\,4A  \citep{Parise2014}. 
In the latter case, the HOOH nondetection disagrees with their prediction, but at the same time, their analysis emphasized the critical role of  the local physical conditions (e.g., density and temperature) and age of the source, neither of which were well determined. 
Their model has therefore not been unequivocally confirmed so far. 

Our aim is to estimate the HOOH abundance during the early phases of YSOs by combining the results of the chemical model from \citet{Du2012} with the assumption of a spherical gradient distribution of the density and temperature of these sources. 
YSOs in their various phases can be very complicated objects geometrically. 
However, for many purposes, the assumption of spherical symmetry with density and temperature gradients that can be determined by spectral energy distribution (SED) analysis is often assumed to be sufficient to obtain good estimates of key parameters of the system \citep{Ivezic1997, Schoeier2002, Kristensen2012}. 
We therefore combined the chemical model from \citet{Du2012}  with simple geometrical assumptions of the YSO in our model. This 'gradient model' will be described elsewhere \citep{Fuchs2020}. 
In brief, because the model assumes spherical symmetry, the YSO is divided into concentric shells, each of which has constant density and temperature. 
The density and temperature of each shell $i$ at radius $r_{shell, i}$ is given by the functions $\rho_{shell, i}\sim {r_{shell,i}}^{-p}$ and  $T_{shell, i}\sim {r_{shell, i}}^{-b}$, with $p$ and $b$ being specific source parameters that have been determined from SED observations.    
In each shell the chemistry starts with the same initial chemical composition, but then evolves differently over time according to the chemical model given by \citet{Du2012}.
The chemical model includes gas-phase and surface reactions, but particular attention was paid to the role of surface reactions, that is, the HOOH reaction pathway, in the water formation process.
In this model, HOOH and  O$_2$ are mainly formed on grain surfaces by hydrogenation of molecular oxygen via
$H+O_2 \rightarrow O_2H,$ with the subsequent reactions $H+O_2H \rightarrow H_2O_2$ and $H+H_2O_2 \rightarrow H_2O + OH$. 
Alternative surface reaction routes are also considered, such as $H + O \rightarrow OH$ and $H+OH \rightarrow H_2O,$ but they do not result in comparable amounts of HOOH or H$_2$O. 
Using our gradient model, we predict the gas-phase abundance of HOOH 
in the close-by YSOs L1551-IRS\,5, R CrA-IRS\,5A, and NGC1333-IRAS\,2A, as well as in the pre-stellar core L1544. We compare our values with the astronomically observed values.

%__________________________________________________________________

\begin{table*}[bth]
\caption{\label{t1}Astronomical parameters of the investigated objects.}
\centering
\begin{tabular}{lccccccl}
\hline\hline
source                    &                              & RA(J2000) &  Dec(J2000)      &  v$_{LRS}$ &  distance & telescope &  reference\\
                               &                              &          $\alpha$         &      $\delta$                    &       (km s$^{-1}$)    &    (pc)      &                 & \\
\hline
R CrA-IRS\,5A                       &  protostar, class 1   & 19 01 48.0 & $-$36 57 21.6  & 5.7 &   130 & APEX 12m &  [1]\\
                                           &                            &                   &                          &        &  170 &                 & [2] \\
NGC1333-IRAS\,2A              &      protostar, class 0                        &                   &                          &      &      220    &                 &  [3]  \\
position 1  &    & 03 28 55.6 & $+$31 14 37.1 & 7.7 &  & APEX 12m & [4]$^{a}$\\
position 2  &   & 03 28 53.7 & $+$31 14 28.0 & 7.7 &  & APEX 12m & \\
position 3  &   & 03 28 56.2 & $+$31 15 00.0 & 7.7 &  & APEX 12m & \\
\hline
L1551-IRS\,5                        &  protostar, class 1  & 04 31 19.10 & $+$18 08 09.0  & 6.2 & 140 &  IRAM 30m &  [5]$^{b}$\\
L1544                                 &  prestellar core     & 05 04 18.50 & $+$25 11 10.0  &  7.2 & 140 & IRAM 30m &  [6] $^{c}$\\
\hline
\hline
\end{tabular}
\tablefoot{

$^{a}$ The use of three different positions for NGC1333-IRAS\,2A is motivated by the fact that during the original HOOH detection toward $\rho$ Oph A, the strongest signal was observed between the SM1 and SM1N region and not at the respective CO peak positions, see \citet{Bergman2011a, Bergman2011b} . 
$^{b}$ For recent source position data of L1551-IRS\,5 using the VLA at 7mm, also see \citet{Villa2017}.
$^{c}$ The L1544 position deviates from the dust peak position $\alpha$: 05 04 17,21; $\delta$: $+$25 10 42,8 and was chosen according to the CH$_3$OH maps from \cite{Bizzocchi2014}. \\
References: [1] \citet{Peterson2011}, 
[2] \citet{Nutter2005}, 
[3] \citet{Bottinelli2007}, 
[4] \citet{Wilkin2002}, 
[5] \citet{Osorio2003},
[6] \citet{Bizzocchi2014}. 
}
\end{table*}

%__________________________________________________________________

\section{Astronomical sources}
\label{sec-astro}
Based on an extensive  list of pre- and protostellar objects (see \citet{Kristensen2012, Mottram2014}), we selected sources that according to our model are the most promising candidates for a successful HOOH detection and that are within reach of either the APEX 12m or the IRAM 30m telescope. These sources are R CrA-IRS\,5A, NGC1333-IRAS\,2A, and L1551-IRS\,5. 
An additional source, the pre-stellar core L1544, was included for the following two reasons. 
First, our model shows that even at early stages, HOOH is expected to be produced. 
Second, L1544 shows water vapor emission in its cold inner core (7-10 K) where water should be frozen out onto grains \citep{Caselli2012}. 
A surprisingly effective desorption mechanism therefore appears to drive the water into the gas phase, and this may also apply to HOOH. 

In all presented sources, water vapor has been observed in emission and for some also in absorption \citep{Kristensen2012}. 
Depending on the method and region (e.g., outer or inner envelope, jet), the reported H$_2$O abundance varies strongly.  
According to Kristensen et al. (2012), column densities of $N_{(H_20)} > 10^{13}$ cm$^{-2}$  are required to account for the absorption features of the ortho-H$_2$O transitions in the outer envelope seen by {\it Herschel}. % mission {\it WISH}.  
Generally, the relative abundances [H$_2$O]/[H$_2$] inferred from absorption features appear to lie between <$10^{-11}$ for highly embedded sources and >$10^{-9}$ for less deeply embedded sources.  
However, NGC13333-IRAS\,2A is an exception with slightly higher relative abundances (see below). 
Investigations of similar sources by comparing CO/H$_2$O line ratios result in water abundances relative to H$_2$ of $10^{-7} - 10^{-5}$ for low-density regions and $10^{-8} - 10^{-6}$ for high-density areas \citep{Franklin2008, Lefloch2010, Kristensen2012}.  
It appears that water emission can be seen mainly in Class 0 objects by outflow components that are caused by in-falling envelopes. 
Class 1 objects reveal only minor or no outflow features, and thus H$_2$O signals are generally weaker.  
Observational details of all four investigated sources are given in Table~\ref{t1}. 
 
{\bf L1544 (prestellar core).}  
This object is a star-less low-mass star-forming region in the constellation Taurus. 
The region is thought of as being on the point of becoming unstable, that is, starting star formation \citep{Crapsi2005, Crapsi2007}.
A mapping of L1544 at 850 $\mu$m is presented in \citet{Shirley2000}.
L1544 can be modeled with a simple flat temperature and density distribution according to \citet{Caselli2002}.
The relative O$_2$ abundance has been shown to be [O$_2$]/[H$_2$] < 6.3$\cdot10^{-8}$ (\cite{Wirstrom2016}).
Water emission has been detected in this source \citep{Caselli2012} with a column density $N_{H_2O}$(ortho+para) > $10^{13}$ cm$^{-2}$ using the ortho-H$_2$O transition $1_{10}-1_{01}$ , which was measured with an integrated line strength of  $\int T_{B}\,dv = 5$ mK km s$^{-1}$. 
This yields a relative water abundance [H$_2$O]/[H$_2$] < $1.4 \times 10^{-10}$ (using $N(H_2) \simeq 7 \times 10^{22}$ cm$^{-2}$). 

{\bf NGC1333-IRAS\,2A (Class 0).}
NGC1333-IRAS\,2A is a typical Class 0 object \citep{Brinch2009} located in the constellation Perseus.  
The source has a complex structure with a clear indication of jet formation \citep{Sandell1994, Wilkin2002, Engargiola1999} and protostellar infall \citep{Ward-Thompson1996}. 
Maps and further information on shocks and the structure at small scales can be found in \citet{Jorgensen2004a, Jorgensen2004b}.
Because the structure is complex, we decided to investigate three spatially separated regions of this source. 
% {GWF: I find the expression ``failed to find'' a bit harsh. What I meant is that the HOOH intensity peak is not located at the position of the continuum intensity peaks, i.e. at the SM1 and SM1N position; but in-between.            
The reasoning behind this is also that \citet{Bergman2011a} did not find the peak intensity of HOOH toward the submillimeter sources SM1 or SM1N within the $\rho$ Oph A complex, but instead found them in between these sources. 
It therefore seemed reasonable to investigate at least two further locations of the IRAS\,2A region that are both differently positioned relative to the existing outflows, that is, position 2 and 3 are still in the vicinity of the central region where densities are still high but either off-center (position 2) or center-line to the north-south outflow.
In the {\it WISH}/{\it Herschel} program, water emission at 557 GHz ($1_{10}-1_{01}$ transition) was found with  $\int T_{mb}^{total} dv = 5.2$ K km s$^{-1}$ \citep{Kristensen2012}. 
The relative abundance was determined as [H$_2$O]/[H$_2$] $\sim 10^{-8}$ \citep{Kristensen2010, Kristensen2012} based on absorption features.  

{\bf {R CrA}-IRS\,5A (Class 1).} 
The source IRS\,5 \citep{Taylor1984} is located in Corona Australis close to the star R CrA and is also known as SMM4, MMS12, or TS2.4 (see A.19 in \citet{Peterson2011}).
% {GWF: I made changes here to be more clear about what is what.
It is a binary system separated by $\sim$78 AU. 
The brighter source IRS\,5a, is of spectral type K5-K7V, and is classified as a Class 1 YSO \citep{Chen1993, Nisini2005}. 
In its direct vicinity (north) lies IRS\,5N. 
SMA continuum measurements at 225 GHz suggest that the latter source is the driver of a CO outflow, whereas IRS\,5a/b shows no features \citep{Peterson2011}. 
Water vapor has also been seen in emission at 557 GHz, with a line intensity of  $\int T_{mb}^{total} dv = 3.8$ K km s$^{-1}$ \citep{Kristensen2012}. 
\citet{Schmalzl2014} reported a relative water vapor abundance of $1.7 \times 10^{-8}$ (using $N_{Hydrogen} \simeq 4.9 \times 10^{22}$ cm$^{-2}$).

{\bf L1551-IRS\,5 (Class 1).}
The close-by YSO L1551 is a Class 1 source with a core binary system in the Taurus molecular cloud complex \citep{Osorio2003, Lee2014, Ainsworth2016}. 
Because of its proximity, luminosity, and accessibility for the IRAM 30m telescope, this source was considered to be a good candidate for our purposes.  
In earlier studies \citep{White2000} no water emission was found toward IRS\,5. 
However, using the Herschel space observatory, H$_2$O at 557 GHz could be detected, and this resulted in an integrated line intensity of $\int T_{mb}^{total} dv = 0.7$ K km s$^{-1}$ \citep{Kristensen2012}.  
The relative gas-phase water abundance is about $4 \times 10^{-10}$ (using $N_{Hydrogen}  \simeq 7.4 \times 10^{23}$ cm$^{-2}$), see \citet{Schmalzl2014}.

%__________________________________________________________________

\begin{table}[tbh]
\caption{\label{obs-tab} Observational details}
\centering
\begin{tabular}{cccc} 
\hline  \hline
  Frequency bands & rms$^{a}$   &  weather  &  $T_{sys}$ \\ 
                           &         &  cond.    &                  \\
  (GHz)                & (mK)  &  $\tau$  &  (K)            \\
\hline
 \multicolumn{4}{l}{APEX 12m, position switching}  \\
 217.000 - 221.000 & 7.8 - 12.8  &   0.071 - 0.147 & 160 -173 \\ 
 249.914 - 253.915 & 23.3          &  0.139             & 208 \\ 
 316.222 - 320.223 & 12.3-47.6  &  0.151 - 0.309  & 240 - 296 \\ 
 \hline
  \multicolumn{4}{l}{IRAM 30m, position switching}  \\
   83.886 -   91.665 & 6.6 & 0.057 & 150 \\ 
 126.878 - 135.159 & 2.0 &  0.081 & 119 \\
 142.558 - 150.838 & 2.1 &  0.081 & 119 \\
 218.070 - 225.850 &45.7 & 0.317 & 702 \\ 
  \hline
  \multicolumn{4}{l}{IRAM 30m, frequency switching}  \\
   87.064 -    94.844 & 2.8 - 3.6 & 0.052 - 0.101  &  140 - 158\\ 
 143.055 - 150.838  & 3.6 - 5.0 & 0.106 - 0.262  &  170 - 232\\
\hline \hline
\end{tabular} 
\tablefoot{
$^{a}$ The rms level depends on the time of integration and was different for each source and frequency band, e.g., the best APEX result was achieved for  NGC1333-IRAS\,2A (position 1) with a 171 min integration time at 219 GHz with an rms of 7.8 mK. The best IRAM rms (= 2.0 mK) level was reached for L1551-IRS\,5 for the HOOH transition at 131.396 GHz with an integration time of 325 min.     
}
\end{table}

\section{Observations}
\label{sec-observations}
The observations were performed using two telescopes: the APEX 12m and IRAM 30m telescopes. We used the APEX 12m telescope in Chile to observe R CrA-IRS\,5A and NGC1333-IRAS\,2A between September 8 and September 19, 2015, during nine observational runs\footnote{APEX 12m project E-096.C-0780A-2015}.  
Observations were made at  217.0-221.0 GHz, 249.9-253.9 GHz, and 316.2-320.2 GHz with a total on-source time of 10 hours; see Table~\ref{obs-tab} for more details. 
The precipitable water vapor (pwv) values were between 0.8\,mm and 2.3\,mm but below 1.3\,mm most of the time. 
We used the HET230 and HET345 (each with 4 GHz bandwidth) heterodyne single-sideband (SBB) front-ends combined with the fast Fourier transform spectrometer XFFTS2 as spectral back-end with a 2500 MHz bandwidth per input (four inputs). 
All APEX measurements were made in position-switching mode, or more precisely, in its variant as wobbler switching (wobbler throw of 60\arcsec), where only the subreflector is moved instead of the entire telescope. 
We checked the pointing and focus every $\sim$1.5 h and $\leq$4 h, respectively.    

The observations using the IRAM 30m telescope at the Pico del Veleta in Spain toward L1551-IRS\,5 and L1544 were made during three consecutive nights in August 2016 (August 24 to 26) \footnote{IRAM 30m project ID: 097-15, run 003-16}. 
On the first night, L1551 was investigated in position-switching mode at 83.9-91.7 GHz, 126.9-135.2 GHz, 142.6-150.8 GHz, and 218.1-225.9 GHz using the heterodyne Eight MIxer Receiver (EMIR) at 3mm (E090), 2mm (E150), and 1.3mm (E230) wavelengths in SSB mode \citep{Carter2012}. 
Two polarizations, vertical and horizontal, were recorded simultaneously. 
As back-end the fast Fourier transform spectrometer (FTS200) with a 200\,kHz resolution and 7.78\,GHz bandwidth per sideband was used\footnote{See "IRAM 30-meter Telescope
Observing Capabilities and Organisation" by Carsten Kramer from July 22, 2016; http://www.iram.fr/GENERAL/calls/w16/30mCapabilities.pdf }.
During the subsequent two nights, L1551 and L1544 were investigated using the frequency-switching mode at 87.1 - 94.8 GHz and 143.1 - 150.8 GHz. 
The weather conditions were favorable during most of the time, as were the overall system conditions; see Table~\ref{obs-tab}. 
The total on-source time at IRAM 30m was 17.6 hours.  

The basic data reduction and processing was made using the continuum and line analysis single-dish software (CLASS) from the GILDAS\footnote{GILDAS is a software provided and maintained by the Institute de Radioastronomie Millim\'etrique (IRAM), see http://www.iram.fr/IRAMFR/GILDAS} software package. As a standard procedure, we checked the quality of the scans and eliminated bad channels. Then we made baseline corrections with a subsequent averaging over individual scans. We identified molecular line positions and used a Gaussian line shape or a superposition of Gaussian line shapes\footnote{Using the CLASS software within the GILDAS package.} to extract their velocity-integrated intensities $W^{*}$ and line widths (full width at half-maximum, FWHM).

Because for both telescopes the output spectra are provided as calibrated antenna temperatures ($T_A^*$), the brightness temperature $T_B$ has to be calculated according to the telescope in use, for example, through the main beam temperature $T_{mb}$ (see \citet{Velilla-Prieto2017} for a similar approach), 
\begin{equation}
T_B = T_{mb} \, \text{\it bff}^{-1} = T^*_A \, \eta^{-1} \, \text{\it bff}^{-1}  \\
\end{equation}
\begin{equation}
\text{with}  \, \, \eta^{-1}= F_{\text{eff}}/B_{\text{eff}} \,\,  \text{and} \,\, \text{\it bff}^{-1} = (\theta_b^2+\theta_s^2)/\theta_s^2 
\label{conv-eq}
.\end{equation}
Here, {\it bff}  is the beam-filling factor,   $B_{\text{eff}}$ is the main-beam efficiency of the antenna, $F_{\text{eff}}$ is the forward efficiency of the antenna, $\theta_s$ is the source size (diameter of the emitting region), and $\theta_b$ is the  half-power beam width (HPBW) of the main beam of the antenna\footnote{The telescope antenna parameters are taken from  http://www.apex-telescope.org/telescope/efficiency/ (values adapted from \citet{Gusten2006}) for the APEX 12m telescope 
and  https://www.iram.es/IRAMES/mainWiki/IRAM30mEfficiencies (values from 26 August 2013 given in the "Improvement of the IRAM 30m telescope beam pattern" report by Carsten Kramer, Juan P\~{e}nalver, and Albert Greve, Version 8.2 ) for the IRAM 30m.}. 
The HPBW $\theta_b$ of the antenna main beam can be approximated for each specific telescope using the expression 
\begin{equation}
\theta_b('')= a / \nu(GHz) 
\label{mb-eq}
,\end{equation}
with $a_{APEX}= 6091.3 [GHz/('')]$ and  $a_{IRAM}= 2446.0 [GHz/('')]$.
The main-beam efficiency $\eta$ can be calculated using the approximation \citep{Velilla-Prieto2017} 
\begin{equation}
\eta^{-1} \equiv F_{\text{eff}}/B_{\text{eff}} =  b \, \exp{(\nu(GHz)/c)^2}
\label{bff-eq}
,\end{equation}
with $b_{\text{APEX}}= 1.233$ and $c_{\text{APEX}}=1074.0 (GHz)$ for the APEX 12m telescope and 
$b_{\text{IRAM}}= 1.115$ and $c_{\text{IRAM}}= 399.5 (GHz)$ for the IRAM 30m telescope.

%__________________________________________________________________

\section{Data analysis methods}
\label{sec-methods}
The observations were analyzed using two methods. 
In method 1 we used the measured integrated intensities $W$ (in [K km s$^{-1}$]), that is, $W^*$ corrected for and calculated using T$_B$ instead of T$_a^*$, to produce a population diagram (also known as Boltzmann plot or rotational diagram) of each detected molecule. 
By assuming local thermodynamic equilibrium (LTE) conditions, we were able to extract the respective total column density $N_C$ and rotational temperature $T_{rot}$ of the molecule under investigation.
% {GWF: changes made as suggested}
The equation we used is similar to the one used in \citet{Goldsmith1999} and \citet{Velilla-Prieto2017} and is given by 
\begin{equation}
\ln\left( \frac{N_u}{g_u}\right) = \ln \left( \frac{3 k_B W}{8 \pi^3 \nu \, S_{ul} \, \mu^2} \right) = \ln \left( \frac{N_C}{Q}\right) - \frac{E_u}{k_B T_{rot}}
\label{Boltzmann-eq}
,\end{equation}
with $N_u$ the column density of the upper level, $g_u$ the degeneracy of the upper level, $k_B$ the Boltzmann constant, $\nu$ the rest frequency of the molecule transition, $S_{ul}$ the line strength, $\mu$ the dipole moment of the molecule, $Q$ the partition function, and $E_u$ the upper energy level of this transition.  
For many astrophysically relevant molecules, the information needed ($\nu$, $S_{ul}$, $\mu$, $Q$ and $E_u$) is provided in databases such as the Cologne Database for Molecular Spectroscopy (CDMS)\footnote{See http://www.astro.uni-koeln.de/cdms/, \citet{Endres2016}.}, or the Jet Propulsion Laboratory (JPL) molecular lines 
catalog \footnote{See https://spec.jpl.nasa.gov/, \citet{Pickett1998}.}. 
References to the original spectroscopic literature can be found there as well.    

%---------- Method 2 -----------
The second method uses the eXtended CASA Line Analysis Software Suite (XCLASS)\footnote{See https://xclass.astro.uni-koeln.de/} by \citet{Moller2017}.  
XCLASS is a supplementing software for the Common Astronomy Software Applications package (CASA) \citep{McMullin2007} 
and also works well with python\footnote{Python Software Foundation, https://www.python.org/} -based software.  
From this software package we mainly used the myXCLASS program to model our data. 
This program  models a spectrum by solving the radiative transfer
equation for an isothermal object \citep{Stahl2005}. 
Similar to the aforementioned method, myXCLASS assumes LTE,  meaning that per molecule only one excitation temperature is assumed for all transitions\footnote{CLASS can separate between several emitting regions (core layers) with different excitations temperatures, velocity offsets, etc.
However, we only used one core layer to analyze the data. }.  
From the input spectrum the program calculates the temperature as well as the column density $N_C$ of the molecule.   
Thus, both methods give quantities that can be directly compared with each other. 

\begin{table}[t]
\caption{\label{obs_results_HOOH} Observational results at HOOH transition frequencies}
\centering
\begin{tabular}{lcccc} 
\hline  \hline
Obs.               &  beam  &  rms   & FWHM$^{a}$   &  $\int T_{mb} dv$  \\ 
frequency      &  size     &          &   (expect.)         &                                \\
  (MHz)           &      (")   & (mK)  & (km $s^{-1}$ )  &  (mK km $s^{-1}$)   \\
\hline 
 \multicolumn{5}{l}{R CrA -IRS\,5A}    \\ 
219 166.8600  &  27.8 &  9.7(4) & 0.7 - 1.4 & <6.8 - 13.6\\
251 914.6794  &  24.2 &  23.7   & 0.7 - 1.4 & <16.6 - 33.2\\
318 222.5200  &  19.1 & 38.6    & 0.7 - 1.4 & <27.0 - 54.0\\
318 712.1000  &  19.1 & 34.5    & 0.7 - 1.4 & <24.2 - 48.3\\
 \multicolumn{5}{l}{NGC1333-IRAS\,2A-1 (center position/position 1)$^*$} \\ 
219 166.8600  & 27.8 &  8.0  & 0.66 - 2.7$^{b}$ &  <5.3 - 21.6 \\
318 222.5200  & 19.1 &  43.2 & same$^{c}$&  <28.5 -116.6 \\
318 712.1000  & 19.1 & 40.6 &  same$^{c}$&  <26.8 - 109.6  \\
\multicolumn{5}{l}{L1551-IRS\,5}    \\
  90 365.5097   &  26.8 & 6.6   & 1.4 - 2.5 &   <9.9 - 16.5 \\
143 712.6282   &  21.7 & 2.1   & 1.4 - 2.5 &   <3.2 - 5.3\\
219 166.8600   &  14.6 &  45.4  & 1.4 - 2.5 &  <63.6 - 113.5  \\
 \multicolumn{5}{l}{L1544}    \\
  90 365.5097   &  26.8 & 2.8   &  0.6 - 1.8  &  <1.7 - 5 \\
143 712.6282   &  21.7 & 5.0   & 0.4 - 0.9  &  <2 - 4.5 \\
\hline \hline
\end{tabular} 
\tablefoot{
$^{a}$ Expected line width inferred from line widths of other molecules (H$_2$CO, CH$_3$OH, etc.) in the same source. 
$^{b}$ Inferred from H$_2$CO line at 218.22 GHz. The HOOH line width is most likely around 1.5  km $s^{-1}$.
$^{c}$ Because no line could be detected in the full spectral range of 318~GHz, the line width was adopted from the 219~GHz value. 
$^{*}$ No detectable line on NGC1333-IRAS\,2A positions 2 and 3. 
}
\end{table}

%__________________________________________________________________
\section{Observational results}
\label{sec-results}

% {GWF: Changes made according to suggestion.}
We focused on the analysis of  HOOH\footnote{The spectroscopic data of HOOH are based on \citet{Petkie1995} and \citet{Camy-Peyret1992}.} 
toward the objects described above, and the observational results are listed in Table~\ref{obs_results_HOOH}.  
Other molecular lines that we observed in these sources in the mentioned frequency ranges are not discussed further. In R CrA-IRS\,5A, for instance, we identified  $^{13}$CO, SO, NO, c-C$_3$H$_2$, CN isotopologs, DCN, and HCCNC.  
The molecules H$_2$CO and CH$_3$OH are discussed in another study, together with detailed information on the model \citep{Fuchs2020}.

%_ _ _ _ _ _ _ _ _ _ _ _ _ _ _ _ _ _ _ _ _ _ _ _ _ _ _ _ _ _ _ _ _ _ _ _ _ _ _ _ _ _ _ _  

  \begin{figure}[t]
   \centering
   \includegraphics[width=\hsize]{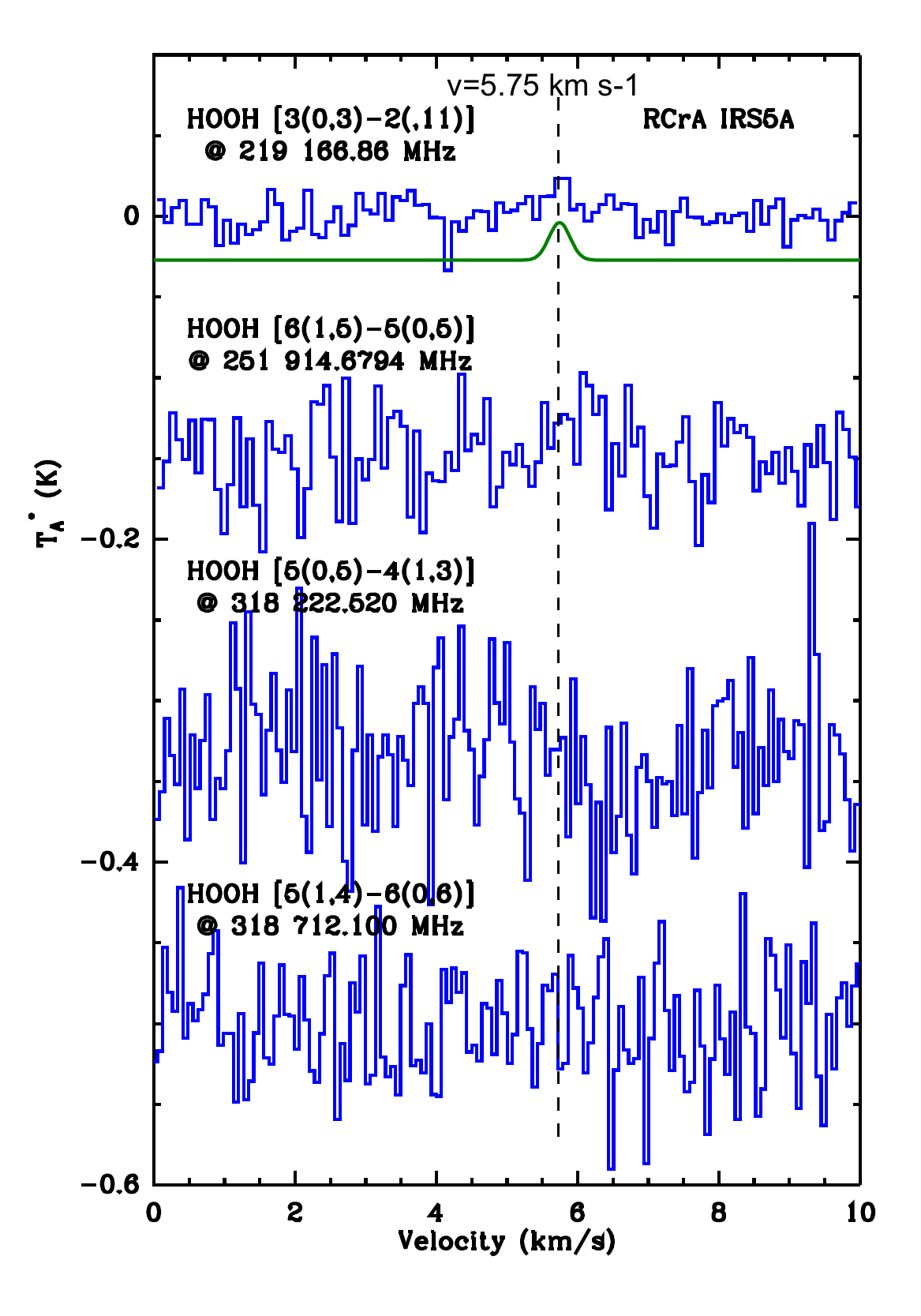}
      \caption{Region with HOOH transitions toward R CrA-IRS\,5A. The green line indicates the HOOH line position. Its line intensity is consistent with the rms level. 
      All spectra have the same intensity scale, but (except for the first) are plotted with an intensity offset. }
         \label{HOOH_RCrA}
   \end{figure}
   
According to the model temperatures in the sources (see Table~\ref{model_obs}), and the specifications of the telescopes, we focused our observations on the  219.2 GHz line for R CrA-IRS\,5A and NGC1333-IRAS\,2A using the APEX 12m telescope   
and the 143.7 GHz line for L1551-IRS\,5 and L1544 using the IRAM 30m telescope. 
In the following paragraphs the observations for each source will be discussed in detail. 
% {GWF: I reformulated the following sentence}
We show below that HOOH is not clearly detected in any of the investigated sources. Instead, 
only upper limits for HOOH column densities can be derived for the individual sources, defined by their respective rms levels.
In  Table~\ref{obs_results_HOOH} a list of all targeted HOOH frequency positions is given, including the integrated line intensities at these positions. 
Table~\ref{HOOH_transitions} in the Appendix shows a list of relevant HOOH transitions, including spectroscopically relevant parameters used in Eq.~\ref{Boltzmann-eq}\, \footnote{$S\mu^2$ can be calculated by converting the 
intensity $I$ [nm$^2$MHz] at a given temperature $T$ using 
\mbox{$S_g \mu_g^{2} = 2.40251\times10^{4} \, I(T) \, Q_{rs}(T) \, \nu^{-1} \, (e^{-E"/kT} - e^{-E'/kT})^{-1}$}, 
see \citet{Pickett1998}. 
The partition function values $Q(T)$ were calculated using the tables given in the CDMS and JPL catalogs.
}. 
 \\
{\bf R CrA-IRS\,5A.} As an example, the R CrA-IRS\,5A observation is shown in Fig.~\ref{HOOH_RCrA}. 
Here, the 219.2 GHz region was observed using 1008 scans, resulting in 150 min integration time and an rms level of 9.7 mK.  
As an example, a simulated line profile with an FWHM of 1.4 km s$^{-1}$ is included at the expected HOOH frequency position close to 219.2 GHz. 
With the analysis methods we described above (Sec.~\ref{sec-methods}), the integrated  line intensity can be used to infer an upper limit of about 6-9$\cdot$10$^{11}$ cm$^{-2}$  (see Table~\ref{model_obs}). 
This result  is not very sensitive to the assumed temperature, that is, between 13 and 25 Kelvin. 
The upper limit is also consistent with other observed frequency regions, as we verified using the myXClass program.  
\\
Spectra at HOOH line positions for the other sources,  that is, for NGC1333-IRAS\,2A,  L1551-IRS\,5, and L1544, can be found in the Appendix section (Fig.~\ref{HOOH_IRAS_2A}, Fig.~\ref{HOOH_L1551} and Fig.~\ref{HOOH_L1544}).  
Below we only summarize the conclusions.
\\
{\bf NGC1333-IRAS 2A.} This source was investigated at three spatial positions (see Table~\ref{t1}). 
The line analysis for this source is not as straightforward as for the calm source close to R CrA. 
For the center position (position 1), the line profile may be different from the assumed Gaussian line shape, as can be observed for the  H$_2$CO lines (not shown here, see \cite{Fuchs2020}), which are more similar to P Cygni profiles. 
Despite the long integration time of 171 minutes for position 1 and 119 minutes for position 3,  HOOH is not detected at any of the frequency settings; see Fig.\ref{HOOH_IRAS_2A}. 
Assuming a temperature of around 22~K results in an upper limit for the HOOH column density toward position 1 of about 2-7$\cdot$10$^{11}$ cm$^{-2}$.
\\
{\bf L1551-IRS\,5.} Although we used a large telescope (30m dish) and very long integration times (325 min in position-switching mode), which allowed us to see an rms level of only 2.1 mK at 143 GHz,  no signal at the HOOH ($4_{1,3}-3_{0,3}$)
frequency position could be detected; see Fig.~\ref{HOOH_L1551}.  
Similar to NGC1333-IRAS 2A, the expected line profile is more complicated because of the influence of a molecular outflow (see H$_2$CO and CH$_3$OH spectra in \citet{Fuchs2020}). 
The source was observed in position- and frequency-switching mode (400 min; 3.6 mK rms level) at 143 GHz. 
Using the model temperature of 21~K, we determined the upper limit for the HOOH column density as about 2$\cdot$10$^{11}$ cm$^{-2}$.
\\
{\bf L1544.} Generally, this source reveals narrow lines that can be easily fit using a Gaussian line profile.  The deconvolved spectra taken in a frequency-switching mode are shown in Fig.~\ref{HOOH_L1544}. 
Assuming a temperature of 12.5~K \citep{Caselli2002}, we determine the upper limit for the HOOH column density as 2$\cdot$10$^{11}$ cm$^{-2}$. 
\\
In conclusion, our study yields a set of upper column density values for HOOH that span 2 - 9$\cdot$10$^{11}$ cm$^{-2}$ for the investigated sources.

%_ _ _ _ _ _ _ _ _ _ _ _ _ _ _ _ _ _ _ _ _ _ _ _ _ _ _ _ _ _ _ _ _ _ _ _ _ _ _ _ _ _ _ _  
\section{Comparison with model data}
\label{sec-comparison}

\begin{table*}[t]
\caption{\label{model_obs} Comparison between the gradient model predictions for HOOH and observational results (rotational diagram method and myXClass)}
\centering
\begin{tabular}{l|ccc|cc|cc} 
\hline  \hline

source   &      \multicolumn{3}{c|}{gradient model}                    & \multicolumn{4}{c}{observational results}  \\  
            &  HOOH      &                             &                              & \multicolumn{2}{c|}{rotational diagram}   &  \multicolumn{2}{c}{myXClass} \\
            &  emission    &    aver. T  $^{*}$   &  N$_T$(HOOH)   $^{*}$    &    T$_{rot}$  &   N$_T$(HOOH) $^{**}$            &    T$_{ex}$    &  N$_T$(HOOH) \\
            &  region       &                             &                              &                   &  upper limits                      &                      &  upper limits \\
            &      (")         &  (K)                      &  (cm$^{-2}$)          &   (K)            &   (cm$^{-2}$)                   &  (K)               &  (cm$^{-2}$)   \\
\hline 
%\multicolumn{8}{l|}{\bf HOOH}         \\
R CrA-IRS\,5A                   &  64.5   &  21(4)             & 5(5)$\cdot 10^{12}$        &  21.4$^{a}$ & $\leq$ 5.8(20)$\cdot 10^{11}$ $^{a}$  &  13-25                &  $\leq$ 8.8(2)$\cdot$10$^{11}$  \\ 
NGC1333-IRAS\,2A-1        &  66.6   &  22(4)             &  $10^{13.2}-10^{16.4}$  &  21.5$^{b}$  & $\leq$ 6.9(56)$\cdot 10^{11}$ $^{b}$  &  22$^{fixed}$     &  $\leq$ 2.7(5)$\cdot 10^{11}$    \\ 
 L1551-IRS\,5                   & 91.8    &  21(4)             & 9(9)$\cdot 10^{12}$         & 21.4$^{c}$ &  $\leq$ 1.6      $\cdot 10^{11}$ $^{c}$  &  17-25                &  $\leq$ 2.2$\cdot 10^{11}$        \\
 L1544                             & 114.3  &  12.5$^{h}$   & $10^{11.7}-10^{14.6}$    & 12.5$^{d}$ &  $\leq$ 1.9(45)$\cdot10^{11}$  $^{d}$  &  12.5$^{fixed}$  &  $\leq$ 1.9$\cdot 10^{11}$        \\
\hline \hline
\end{tabular} 
\tablefoot{$^{*}$ The indicated uncertainties are mainly due to age uncertainties of the sources and the boundary conditions of the chosen integration limits (inner and outer shell radii).  
Early sources such as NGC1333-IRAS\,2A-1   and  L1544  show a high age dependence for HOOH.   
$^{**}$Column densities were obtained by applying the rotational diagram technique. 
For HOOH, the observational upper limit is determined using the average and broader line width values  from Table~\ref{obs_results_HOOH}.\\
$^{a}$ A temperature of 21.4~K (model value) was assumed. However, the value of the column density is not very temperature sensitive, i.e., assuming T=25~K results in nearly the same N$_T$. \\
$^{b}$ Assuming an expected  line width of 1.5 km s$^{-1}$. The 219 GHz rms level was used. Fitted assuming T=21.5~K (model value).  \\
$^{c}$ Assuming a temperature at 21.4~K (model value). The 143 GHz rms was used (because of the longest integration time).   \\ 
$^{d}$ Assuming a temperature at 12.5~K \citep{Tafalla1998}  and using the 143 GHz rms level. \\ 
}
\end{table*}

The results of the observations are summarized and compared to the predicted values by our gradient model in Table~\ref{model_obs}. 
The modeled water vapor results are not mentioned in this table because we do not have any observations of H$_2$O transitions, but the water abundances are discussed below and are compared with results from other groups.  
In none of the four investigated sources were we able to identify hydrogen peroxide  signals, and the estimated column densities are 
expected to be below the derived upper limit values that roughly cover 2-9$\cdot$10$^{11}$ cm$^{-2}$. 
These nondetections of HOOH in protostellar environments are unfortunate, but agree with a number of previous studies in which this molecule could not be detected either. 
     \\
{\bf R CrA-IRS\,5A (Class 1).}
This source proved to be well suited for our applied gradient model. 
The spectra are not dominated by internal source dynamics, such as outflows, and allow a straightforward analysis. 
With a relative  [HOOH]/[H$_2$] abundance of 5$\cdot10^{-10}$ , the model predicted an HOOH abundance nearly an order of magnitude higher than what we derived as upper limit from the observations.     
Geometry aspects, that is, the assumed density and temperature gradient of the YSO, are most likely not the cause for this discrepancy \citep{Fuchs2020}. 
Our modeled gas-phase water abundance is $X$=[H$_2$O]/[H$_2$] = 2$\cdot$10$^{-8}$ and agrees with the $X\approx 10^{-8}$ deduced by \citet{Schmalzl2014}.   
In our model the solid HOOH abundance is indicative for the H$_2$O production and seems to fit to the latter number. 
However, the calculated gas-phase HOOH abundance is clearly too high. 
Although we see no fundamental contradiction between a good agreement of the model and observational values for gas-phase H$_2$O, 
the nondetection of HOOH is therefore still surprising in view of the fact that  H$_2$O abundances (and the abundance of other molecules\footnote{See H$_2$CO and CH$_3$OH abundances in \citet{Fuchs2020}.}) can be well described by the model.
\\
{\bf NGC1333-IRAS\,2A (Class 0).} 
% {GWF: Made change here beacuse of too many ``others''}
Because of the inner structure and dynamics, the deviation between the modeled structure of the object and the observed geometry is stronger than in the other investigated sources. 
To compare the observations with the model, certain assumptions (such as the restriction to the $v_{lsr}=7.34$ km s$^{-1}$ component of the multipeak molecular transition lines) had to be made.
Furthermore, this source is very young, which causes the model to produce unrealistically high HOOH column densities.
Consequently, the predicted HOOH abundances  ([HOOH]/[H$_2$] $\approx$ 7$\cdot$10$^{-10}$ - 10$^{-6}$) could not be confirmed and only an upper limit could be given. 
In case of water vapor, our model results in a relative abundance of [H$_2$O]/[H$_2$] $\approx$3$\cdot10^{-8}$ - 6$\cdot 10^{-7}$ depending on the age of  the source, with the lower value corresponding to a more evolved  source ($\approx$ 10$^5$ yr). 
Observations by \citet{Kristensen2012} indicate a relative water abundance of $\approx$ $10^{-8}$. 
However, according to the same authors, the source is a young Class 0 object ($\sim 2\times10^4$ yrs).  
No conclusive result can therefore be drawn, but given the long integration times at the APEX 12m telescope, further single-dish searches for this molecule toward IRAS\,2A will most likely not result in a detection.     
\\
{\bf L1551-IRS\,5 (Class 1).}
Similar to NGC1333-IRAS\,2A, the spectra reveal dynamical and most likely non-LTE processes within this source. 
In our analysis we focused on the narrow $v_{lsr}=7.0$ km s$^{-1}$ component of the lines.
In our model the column density of water vapor is 3.4$\cdot10^{14}$ cm$^{-2}$ and very close to the observed 3.1$\cdot10^{14}$ cm$^{-2}$ by \citet{Schmalzl2014} \footnote{We used a $N_{hydrogen}$ column density of 1.2$\cdot10^{22}$ cm$^{-2}$ , whereas \cite{Schmalzl2014} used N$_H$ = 7.4$\cdot10^{23}$ cm$^{-2}$ , resulting in a relative H$_2$O abundance of 4$\cdot10^{-10}$. }.
Concerning the HOOH molecule, the column density of about $1\cdot10^{13}$ cm$^{-2}$ predicted by our model could not be confirmed by the observations. 
The corresponding relative [HOOH]/[H$_2$] abundance is $3\cdot10^{-8}$. 
The new observationally determined upper limit is about 40 times lower than this value when a temperature between 17 and 25~K is assumed.    
Like in the other Class 1 source, that is, R CrA-IRS\,5A, the model overestimates the HOOH abundance but produces fairly good predictions for other investigated molecules.     
\\
{\bf L1544 (prestellar core).} Previous work by \citet{Tafalla1998} suggested that molecules can be seen in this source at low temperatures. 
In our model we assumed a 12.5~K temperature for HOOH. 
The upper abundance limit of this molecule is lower than the predicted value (even below the lower value given in Table~\ref{model_obs}),
which is indicative of the already known weakness of the model, which tends to overestimate HOOH at early stages. 
The calculated relative [HOOH]/[H$_2$] abundance is 2$\cdot10^{-12}$ - 2$\cdot10^{-9}$ depending on the age of the source. 
When the relative water vapor abundance is calculated, our gradient model results ($X_{H_2O}$ =  10$^{-8}$ to 10$^{-9}$) disagree with estimated abundances from observations (< 1.4$\cdot10^{-10}$) by \citet{Caselli2012}.   
Because of the low temperatures, \citet{Caselli2012} were surprised to observe water vapor emission in its inner core. 
Our hope that this somewhat unexpected detection might be based on mechanisms that also drive HOOH into the gas phase has not been fulfilled. 

%______________________________________________________________
\section{Discussion}
\label{sec-discussion}

\begin{table}
\caption{\label{target_sources}  List of sources in which HOOH has been searched for.}
\centering
\begin{tabular}{lccc} 
\hline  \hline
 Source  & HOOH     & HOOH                   & Ref.\\ 
              & rms         &  $\int T_{mb} d\nu$    & \\
              & (mK)        & (mK km s$^{-1}$)     & \\
\hline 
$\rho$ Oph-A              &         & 167$^{a}$  & [1] \\
Orion ``H$_2$-Peak 1''  &   26.1$^{b}$    &    172$^{b}$       &  [2]   \\
\hline
r Oph-B2-MM8            & 22.9  &  &   [3]  \\
G15.01-0.67               & 19.2  & &   [3]  \\ 
G018.82-00.28MM1    & 21.1 &  &   [3]  \\
G018.82-00.28MM4    & 21.0 &  &   [3]  \\
G028.53-00.25MM1    & 17.7  & &   [3]  \\
NGC6334I(N)              &  18.6 & &   [3]  \\
G1.6-0.025                 &  16.1 & &   [3]  \\
NGC1333-IRAS\,4A       &  24.6 & &   [3]  \\
L1527                          &  18.4 & &   [3]  \\
R CrA-IRS\,7B                &  23.1 & &   [3]  \\ 
\hline
IRAS\,16293-2422        & <67  & & [4] \\
 \hline
 R CrA-IRS\,5A              &  9.7$^{a}$  & $ \leq$10.2$^{a}$   & [5]  \\
 NGC1333-IRAS\,2A      &  8.0$^{a}$  &  $ \leq$13.5$^{a}$   & [5]  \\
 L1551-IRS\,5               &  2.1$^{c}$   & $ \leq$4.3$^{c}$   & [5]  \\ 
 L1544                         &  5.0$^{c}$   & $ \leq$4.5$^{c,d}$   & [5]  \\
\hline \hline
\end{tabular} 
\tablefoot{
$^{a}$ Transition $3_{0,3}-2_{1,1}$ @ 219.167 GHz;
$^{b}$ Value from transition $3_{0,3}-2_{1,1}$ at 219.167 GHz, which is one of five observed HOOH transitions, with the best S/N ratio. 
Here the integrated area is given in $\int T_{A} d\nu$ and not in $\int T_{mb} d\nu$;
$^{c}$ Value from transition $4_{1,3}-3_{0,3}$ at 143.712 GHz; 
$^{d}$ Instead of using the mean FWHM value in Table~\ref{obs_results_HOOH}, we assumed the higher FWHM value to estimate the upper limit. 
References: [1] \citet{Bergman2011a},
[2] \citet{Liseau2015},
[3] \citet{Parise2014},
[4] TIMASS survey \citet{Caux2011}, and 
[5] {\it \textup{this work}}. 
}
\end{table}

So far, HOOH has been identified in only two sources, one of which (OMC-1) is still not fully verified; see Table~\ref{target_sources}.  
Coincidentally, these sources are the only ones in which the precursor molecule oxygen (O$_2$) could be identified
\footnote{See \citet{Larsson2007, Liseau2012, Goldsmith2011} and \citet{Chen2014}.}, but with very low O$_2$/H$_2$ ratios, for example, O$_2$/H$_2$ $\approx \, 5\cdot10^{-8}$ in $\rho$ Oph A  \citep{Liseau2012} . 

Both sources are in regions where nearby stars cause high irradiation.   
New investigations of the radiation field at $\rho$ Oph A \citep{Lindberg2017} indicate, for example, that the H$_2$CO rotational temperatures in the protostellar envelopes of this region are
strongly enhanced close to the Herbig Be star S1 (similarly, see also the study by \citet{Lindberg2015} for the CrA star-forming region). 
% {GWF: I rephrased the sentence by being more specific who the authors are.} 
\citet{Lindberg2017} write that ``for some reason, the H$_2$CO gas is more prone to be heated by external radiation fields, or the H$_2$CO abundance is enhanced in the irradiated gas, possibly due to photochemistry''. 
The role of an external radiation field has yet not been properly investigated for HOOH. 
For example, it might well be that the chemistry is changed or that desorption is enhanced. 
In the case of water, it is known that photodesorption takes place, but that dissociative reactions on the surface also take place, that is, the transition grain-gas can be destructive and thus decrease the water gas-phase abundance \citep{Andersson2008, Oberg2009}. 

Our model can be applied to a flat geometry and reproduces the HOOH abundance toward $\rho$ Oph A without explicitly taking these effects into account 
(e.g., the desorption rate results mainly from the exothermal energy during the HOOH formation, not from radiation). 
This might be due to these  as yet unknown or very difficult to model effects that are missing or misinterpreted in our chemical model and lead to incorrect results for the sources we investigated here. 
To further investigate the interplay of radiation with HOOH formation and photodesorption, we are currently bound to the two sources $\rho$ Oph A and Orion A.
However, a model refinement can only be successful if more data are available, either in the form of higher spacial resolution of the molecule distribution of HOOH and related grain-borne molecules, or as an increased  set of observed molecular species to obtain a better picture of the local radiation fields 
\footnote{With the recent work by \citet{Larsson2017}, more data on $\rho$ Oph A are now available to further investigate these questions.}. 

Evidence of HOOH in one or two sources, and the nondetection of HOOH in many investigated objects, may be interesting in connection with the chemistry of our own Solar System. 
Has our primordial solar environment been special? 
Measurements by the double-focusing mass spectrometer (DFMS) of the Rosetta Orbiter Spectrometer for Ion and Neutral Analysis (ROSINA) of the coma of comet 67P 67P/Churyumov–Gerasimenko \citep{Bieler2015} indicate abundant molecular oxygen and reveal an hydrogen peroxide to oxygen ratio of H$_2$O$_2$/O$_2$ = (0.6 $\pm$ 0.07) $\cdot 10^{-3}$ which is very close to the ratio measured in the $\rho$ Ophiuchi dense core HO$_2$/O$_2$<H$_2$O$_2$/O$_2$<0.6$\cdot10^{-3}$ \citep{Bergman2011a,Parise2012}.
\cite{Bieler2015} argued that if these gas-phase abundance ratios reflect those in the cometary ice, it would support the existence
of primordial O$_2$. 
% {GWF: changed temperatures in temperatrure} 
Furthermore, the $\rho$ Ophiuchi A core has been suggested to have a higher temperature, that is, about 20–30K, 
compared to 10~K for most other dense interstellar clouds, which is also typical of estimates for the comet-forming conditions
in the outer early solar nebula. This would suggest that our Solar System might have formed from an unusually warm molecular
cloud.
% {GWF: I changed $H_2O_2$ in HOOH} 
However, \citet{Luspay-Kuti2018} pointed out that O$_2$ condensation in the protosolar nebula might not be the only cause of the measured abundances, but that also {\it \textup{in situ}} O$_2$ formation processes (e.g., post-accretion radiolysis) might be at play that connect this molecule with H$_2$O and related molecules such as hydrogen peroxide (e.g., O$_2$ formation via HOOH dismutation or disproportionation).

%______________________________________________________________
\section{Conclusions}
\label{sec-conclusions}

At the heart of this work is the attempt to answer the question whether it can be shown that water is formed on grain surfaces through a chemical pathway that includes hydrogen peroxide as the main intermediate. 
For this we searched for HOOH gas-phase signatures exclusively in the dense environments of YSOs and in a prestellar core and assumed that HOOH can be released into the gas phase, thermally or through other means, after it has formed in the solid state. 
A model that combines a previously used chemical model \citep{Du2012} and a physical model that takes into account the shell-like structure of typical protostellar objects  was used for this purpose (see 'gradient model' in \citet{Fuchs2020}). 
Based on predictions by this model, four sources were chosen in which an HOOH detection seemed feasible and in which water has previously been detected. Despite long integration times on the APEX~12m and IRAM~30m telescopes, we cannot report an unambiguous hydrogen peroxide detection for any of these targets. 
Further support for an H$_2$O solid-state formation scheme through HOOH, that is, by comparing relative abundances, is therefore not possible. 
However, this nondetection does not exclude the proposed solid-state formation scheme. 
As said in the introduction, as an intermediate, HOOH may fully react to water, resulting in rather low ice and also gas-phase HOOH abundances. 
It is also possible that reactive desorption as a possible sublimation process is not efficient enough to build up substantial HOOH gas-phase abundances, and it is equally well possible that HOOH, after it is released from the ice, forms the starting point in a gas-phase reaction. 
At this stage, it is not clear how other not yet included chemical reaction mechanisms, both in the solid state and in the gas phase, will affect the abundance of HOOH. 
For this, follow-up studies are required. 
Another promising development that helps to investigate the surface-reaction scheme would be a direct solid-state detection of HOOH in the infrared region, similar to those available for H$_2$O ice, or other ice-borne molecules such as H$_2$CO and CH$_3$OH. 
This will not be straightforward because only a broad HOOH band around 1400-1500 cm$^{-1}$ is clearly different from H$_2$O vibrational modes. 
Unfortunately, vibrational modes from other abundant species can also be found in this region.
Our hope is that new insights will arise from dedicated ice surveys performed by the James Webb Space Telescope (JWST) in the near future.

%______________________________________________________________

\begin{acknowledgements}
     We thank the APEX 12m and IRAM 30m staff for their excellent support. We thank Peter Schilke, Thomas Möller and Álvaro Sánchez-Monge for their kind intruduction to myXCLASS. 
\end{acknowledgements}

%-------------------------------------------------------------------

%###################################################################################################
%
%       APPENDIX
%
%###################################################################################################
\appendix
\section{Tables and figures}
      
 \begin{table*}
\caption{\label{HOOH_transitions}Relevant transitions of HOOH in the (80 - 700 GHz) submillimeter region}
\centering
\begin{tabular}{ccrccc r@{=}l c} 
\hline \hline
  \multicolumn{2}{c}{Transition}                    &Frequency  & $\log$(Intensity) & $A_{ul}$    & E$_{up}$   &  \multicolumn{2}{c}{upper state}   & spin \\
   rotational            &  tunneling$^{\dagger}$  &                 & at 300K             &                   &                  &  \multicolumn{2}{c}{ degeneracy} & weight\\
($J_{K_a,K_c}-J'_{K'_a,K'_c}$)  & $\tau'-\tau''$ &(MHz)        & (nm$^2$MHz)    & ($s^{-1}$)  & (K)            &  \multicolumn{2}{c}{$g_{up} = g_I \cdot g_k$} & \\ 
\hline
\rule[-1ex]{0pt}{2.5ex} $3_{1,2}-2_{0,2}$ & 3-1 & *90 365.5097   & -4.2979  & 1.213$\times10^{-5}$ & 28.2 &    7& $1 \cdot 7$  & 1\\
\rule[-1ex]{0pt}{2.5ex} $4_{1,3}-3_{0,3}$ & 3-1 & *143 712.6282 & -3.3341  & 4.736$\times10^{-5}$ & 38.1 &  21& $3 \cdot 7$ & 3\\
\rule[-1ex]{0pt}{2.5ex} $5_{1,4}-4_{0,4}$ & 2-4 & 197 561.2819   & -3.4728  & 1.205$\times10^{-4}$ & 50.6 &  11& $1 \cdot 11$ & 1\\
\rule[-1ex]{0pt}{2.5ex} $3_{0,3}-2_{1,1}$ & 4-2 & *219 166.8600 & -3.3556  & 8.584$\times10^{-5}$ & 31.2 &  21& $3 \cdot 7$  & 3\\ 
\rule[-1ex]{0pt}{2.5ex} $6_{1,5}-5_{0,5}$ & 2-4 & *251 914.6794 & -2.7387  & 2.458$\times10^{-4}$ & 65.5 &  39& $3 \cdot 13$ & 3\\ 
\rule[-1ex]{0pt}{2.5ex} $4_{0,4}-3_{1,2}$ & 4-2 & 268 961.1700  & -3.4931   & 1.843$\times10^{-4}$ & 41.1 &    9& $1 \cdot 9$  & 1\\ 
\rule[-1ex]{0pt}{2.5ex} $6_{1,5}-7_{0,7}$ & 3-1 & 270 610.1000  & -2.7781   & 2.544$\times10^{-4}$ & 81.9 &  39& $3 \cdot 13$  & 3\\ 
\rule[-1ex]{0pt}{2.5ex} $5_{0,5}-4_{1,3}$ & 4-2 & *318 222.5200 & -2.7632  & 3.314$\times10^{-4}$ & 53.4 &  33& $3 \cdot 11$ & 3\\ 
\rule[-1ex]{0pt}{2.5ex} $5_{1,4}-6_{0,6}$ & 3-1 & *318 712.1000 & -3.1657  & 4.124$\times10^{-4}$ & 67.0 &  11& $1 \cdot 11$ & 1\\ 
\rule[-1ex]{0pt}{2.5ex} $3_{1,3}-3_{0,3}$ & 2-4 & 616 141.4500  & -1.9131   & 6.762$\times10^{-3}$ & 44.4 &  21& $3 \cdot 7$ & 3\\ 
\rule[-1ex]{0pt}{2.5ex} $2_{1,2}-2_{0,2}$ & 2-4 & 617 459.1800  & -2.5250   & 6.788$\times10^{-3}$ & 37.0 &    5& $1 \cdot 5$ & 1\\ 
\rule[-1ex]{0pt}{2.5ex} $1_{1,1}-1_{0,1}$ & 2-4 & 618 341.7600  & -2.2622   & 6.804$\times10^{-3}$ & 32.2 &    9& $3 \cdot 3$ & 3\\ 
\rule[-1ex]{0pt}{2.5ex} $1_{1,0}-0_{0,0}$ & 3-1 & 670 595.8200  & -2.8432   & 5.786$\times10^{-3}$ & 32.2 &    3& $1 \cdot 3$ & 1\\ 
\hline \hline
\end{tabular} 
\tablefoot{All values are taken from the JPL catalog; \cite{Pickett1998} (entry: 34004 H2O2). Frequencies marked with an asterisk are transitions that have been observationally investigated. $g_I$ is the spin-statistical weight, and $g_k$ is the upper state spin-rotational degeneracy  (2J+1). $^{\dagger}$ Tunneling splitting component \citep{Hunt1965}. 
}
\end{table*}

%=================================================================

% R CrA-IRS 5A
% see main text

%   NGC1333-IRAS 2A

 \begin{figure*}
   \centering
   \includegraphics[width=\hsize]{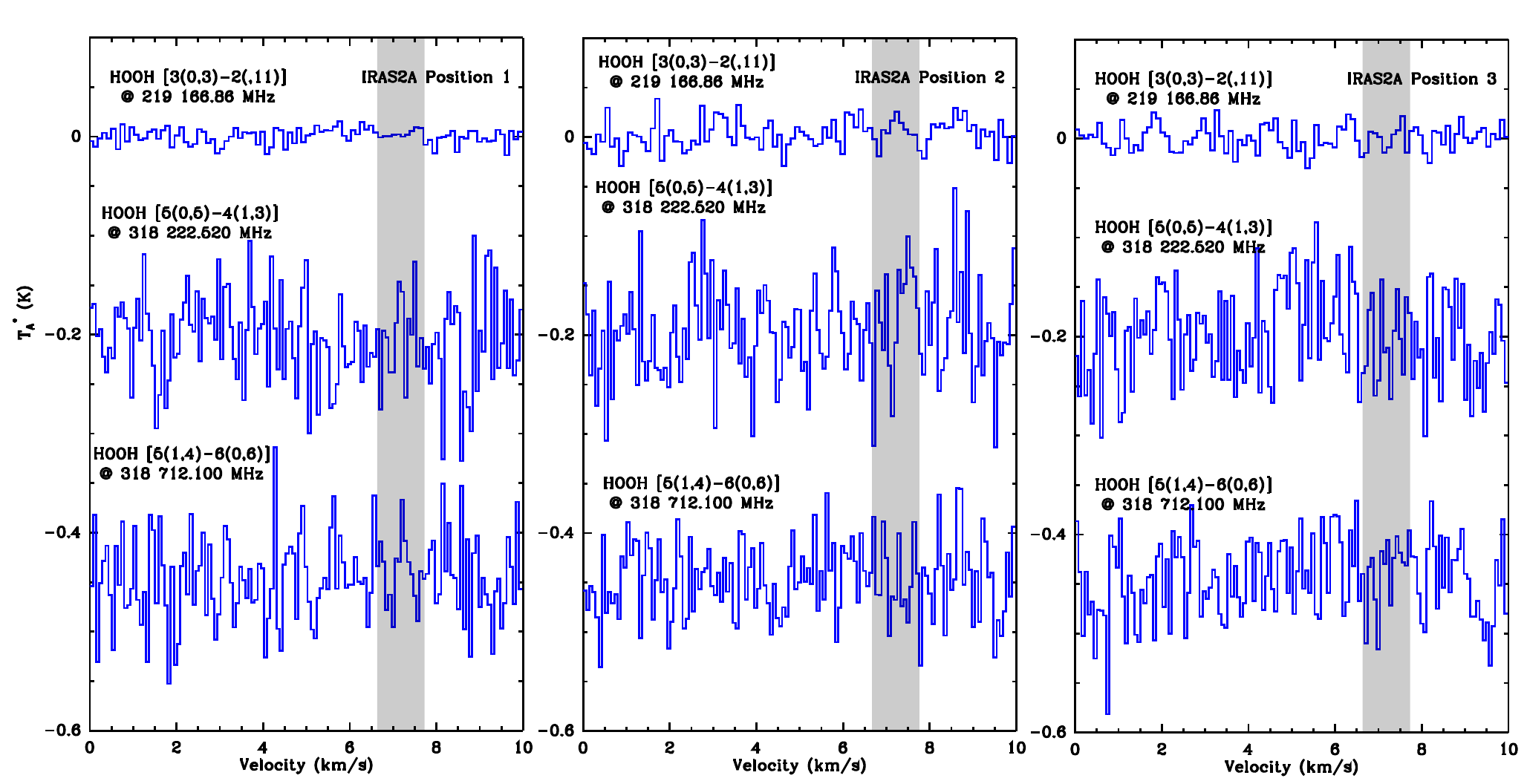}
      \caption{Frequency region with HOOH transitions (gray) toward IRAS\,2A positions 1-3.
          All spectra have the same intensity scale, but (except for the top spectra) are plotted with an intensity offset. 
          }
         \label{HOOH_IRAS_2A}
   \end{figure*}

%   L1551

 \begin{figure*}
   \centering
   \includegraphics[width=\hsize]{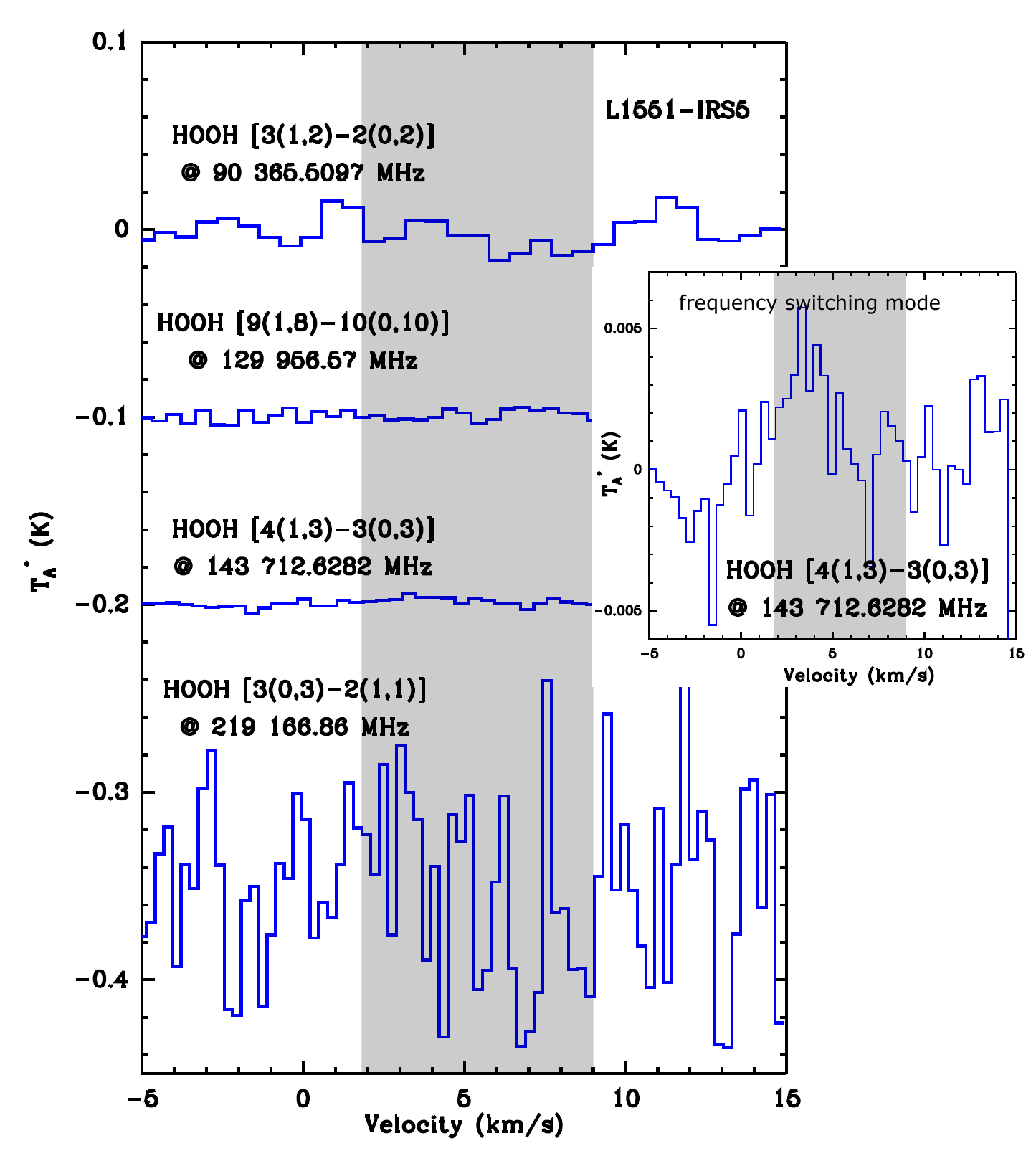}
      \caption{Frequency region with HOOH transitions (gray) toward L1551-IRS\,5 using position-switching mode. 
      All spectra have the same intensity scale, but (except for the first spectrum) are plotted with an intensity offset. 
      The inset shows the same region around the HOOH transition $4_{1,3}-3_{0,3}$ at 143.71263 GHz using frequency-switching mode.} 
         \label{HOOH_L1551}
   \end{figure*}

   % L1544

   \begin{figure}
   \centering
   \includegraphics[width=\hsize]{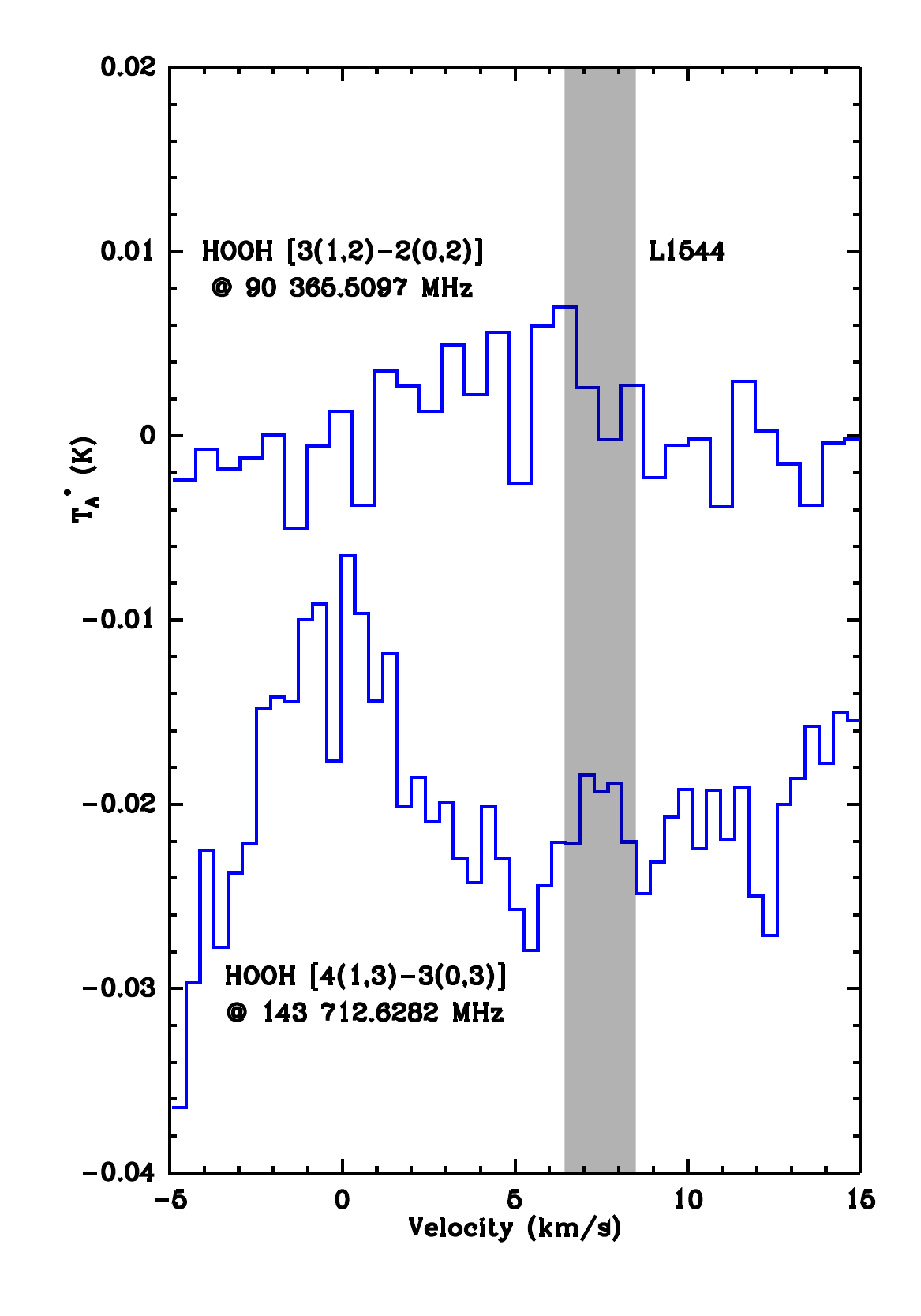}
      \caption{Frequency region with HOOH transitions (gray) toward L1544 (frequency-switching mode). 
                   Both spectra have the same intensity scale, but the lower spectrum is plotted with an intensity offset. }
                 \label{HOOH_L1544}
\end{figure}


\begin{thebibliography}{}



\bibitem[Ainsworth et al. (2016)]{Ainsworth2016}
Ainsworth, R.E., Coughlan, C.P., Green, D.A., Scaife, A.M.M., \& Ray, T.P. 
%, A GMRT survey of regions towards the Taurus molecular cloud at 323 and 608 MHz, 
\ 2016, \mnras, {462}, 2904 %(2016)

\bibitem[Anderson \& van Dishoeck (2008)]{Andersson2008}
%S. Andersson and E. F. van Dishoeck,
Andersson, S. , \& van Dishoeck, E. F. 
%, Photodesorption of water ice - A molecular dynamics study,
\ 2008, \aap, {491}, 907 %(2008)

\bibitem[Bergman et al. (2011a)] {Bergman2011a}
%P. Bergman, B. Parise, R. Liseau, B. Larsson, H. Olofsson, K. M. Menten and R. Güsten,
Bergman, P. , Parise, B. , Liseau, R. , et al. 
%, Detection of interstellar hydrogen peroxide, 
\ 2011, \aap, {531}, L8 %(2011)

\bibitem[Bergman et al. (2011b)] {Bergman2011b}
%P. Bergman, B. Parise, R. Liseau, and B. Larsson
Bergman, P., Parise, B., Liseau, R., \& Larsson, B. 
%, Deuterated formaldehyde in $\rho$ Ophiuchi A, 
\ 2011, \aap, {527}, A39 %(2011)

\bibitem[Bieler et al. (2015)]{Bieler2015}
%A. Bieler, K. Altwegg, H. Balsiger, A. Bar-Nun, J.-J. Berthelier, P. Bochsler, C. Briois, U. Calmonte, M. Combi, J. De
%Keyser, E.F. van Dishoeck, B. Fiethe, S.A. Fuselier, S. Gasc, T.I. Gombosi, K.C.Hansen, M. Hässig, A. Jäckel, E. Kopp,
%A. Korth, L. LeRoy, U. Mall, R. Maggiolo, B. Marty, O. Mousis, T. Owen, H. R$\grave{e}$me, M. Rubin, T. S$\acute{e}$mon, C.-Y. Tzou,
%J.H. Waite, C. Walsh and P. Wurz
Bieler, A. , Altwegg, K. , Balsiger, H. , et al.
%, Abundant molecular oxygen in the coma of comet 67P/Churyumov–Gerasimenko,
\ 2015, \nat, {526}, 678 % (2015)

 \bibitem[Bizzocchi et al. (2014)]{Bizzocchi2014}
%L. Bizzocchi, P. Caselli, S. Spezzano, and E. Leonardo,
Bizzocchi, L. , Caselli, P. , Spezzano, S. , \& Leonardo, E. 
%, Deuterated methanol in the pre-stellar core L1544*, 
\ 2014,  \aap, {569}, A27 %(2014)

 \bibitem[Bottinelli et al. (2007)]{Bottinelli2007}
%S. Bottinelli, C. Ceccarelli, J. P. Williams, and B. Lefloch,
Bottinelli, S., Ceccarelli, C., Williams, J. P., \& Lefloch, B. 
%, Hot corinos in NGC 1333-IRAS4B and IRAS2A,
\ 2007, \aap, 463, 601 %(2007)

 \bibitem[Boudin et al. (1998)]{Boudin1998}
%N. Boudin, W.A. Schutte, and J.M. Greenberg, 
Boudin, N. , Schutte, W.A. , \& Greenberg,  J.M. 
%, Constraints on the abundances of various molecules in interstellar ice: laboratory studies and astrophysical implications
\ 1998, \aap, 331, 749 %(1998)

\bibitem[Brinch et al. (2009)]{Brinch2009}
%C. Brinch, J.K. Jorgensen, M.R. Hogerheijde, 
Brinch, C., Jorgensen, J.K., \& Hogerheijde, M.R. 
%, The kinematics of NGC 1333-IRAS2A – a true Class 0 protostar,
\ 2009, \aap, {502}, 199 %(2009)

\bibitem[Camy-Peyret et al. (1992)]{Camy-Peyret1992}
%C. Camy-Peyret, J.-M. Flaud, J. W. C. Johns and M. No$\ddot{\text{e}}$l
Camy-Peyret, C., Flaud, J.-M., Johns, J. W. C., \& No$\ddot{\text{e}}$l, M. 
%, Torsion-vibration interaction in H$_2$O$_2$: First high-resolution observation of $\nu_3$,
\ 1992, J. Mol.  Spect., {155}, 84 %(1992)

\bibitem[Carter et al. (2012)]{Carter2012}
%M. Carter, B. Lazareff, D. Maier, J.-Y. Chenu, A.-L. Fontana, Y. Bortolotti, C. Boucher, A. Navarrini,
%S. Blanchet, A. Greve, D. John, C. Kramer, F. Morel, S. Navarro, J. Peñalver, K. F. Schuster, and C. Thum
Carter, M., Lazareff, B., Maier, D., et al. 
%, The EMIR multi-band mm-wave receiver for the IRAM 30-m telescope, 
\ 2012, \aap, {538}, A89 % (2012) Astronomy \& Astrophysics,

\bibitem[Caselli et al. (2002)] {Caselli2002}
%P. Caselli, C. M. Walmsley, A. Zucconi, M. Tafalla, L. Dore, and P. C. Myers
Caselli, P., Walmsley, C. M., Zucconi, A., et al. 
%, Molecular ions in L1544. II. The ionization degree,
\ 2002, \apj, {565}, 344 % (2002) Astrophys.J. 


\bibitem[Caselli et al. (2012)] {Caselli2012}
%Paola Caselli, Eric Keto, Edwin A. Bergin, Mario Tafalla, Yuri Aikawa, Thomas Douglas, Laurent Pagani, Umut A. Yíldíz, Floris F. S. van der Tak, C. Malcolm Walmsley, Claudio Codella, Brunella Nisini, Lars E. Kristensen, and Ewine F. van Dishoeck
Caselli, P., Keto, E., Bergin, E.A., et al. 
%, First detection of water vapor in a pre-stellar core, 
\ 2012, \apjl, {759}, L37 %(2012) Astrophys. J. Letters,


\bibitem[Caux et al. (2011)]{Caux2011}
%E. Caux, C. Kahane, A. Castets, A. Coutens, C. Ceccarelli, A. Bacmann, S. Bisschop, S. Bottinelli, C. Comito, F. P. Helmich, B. Lefloch, B. Parise, P. Schilke, A. G. G. M. Tielens, E. van Dishoeck, C. Vastel, V. Wakelam, and A. Walters
Caux, E., Kahane, C., Castets, A.,  et al. 
%, TIMASSS: the IRAS 16293-2422 millimeter and submillimeter spectral survey - I. Observations, calibration, and analysis of the line kinematics, 
\ 2011, \aap, {532}, A23 % (2011) Astronomy \& Astrophysics,

\bibitem[Cazaux et al. (2016)]{Cazaux2016}
%S. Cazaux, M. Minissale, F. Dulieu, and S. Hocuk
Cazaux, S., Minissale, M., Dulieu, F.,  \& Hocuk, S. 
%, Dust as interstellar catalyst II. How chemical desorption impacts the gas, 
\ 2016, \aap, {585}, A55 %(2016) Astronomy \& Astrophysics,
% DOI: 10.1051/0004-6361/201527187


\bibitem[Chen \& Graham (1993)]{Chen1993}    
%Chen, W. P., \& Graham, J. A.
Chen, W. P., \& Graham, J. A.
%, Ice Grains in the Corona Australis Molecular Cloud, 
\ 1993, \apj, {409}, 319 %(1993) Astrophys. J.,
    
\bibitem[Chen et al. (2014)]{Chen2014}
%Jo-Hsin Chen, Paul F. Goldsmith, Serena Viti, Ronald Snell, Dariusz C. Lis, Arnold Benz, Edwin Bergin, John Black, Paola Caselli, Pierre Encrenaz {\em et al.}
Chen, J.-H., Goldsmith, P.F., Viti, S., et al. 
%, Herschel HIFI Observations of O$_2$ Towards Orion: Special Conditions for Shock Enhanced Emission,
%%HERSCHEL HIFI OBSERVATIONS OF O2 TOWARD ORION: SPECIAL CONDITIONS FOR SHOCK ENHANCED EMISSION, 
\ 2014, \apj, {793}, 111 %(2014)  Astrophys.J., 

\bibitem[Cuppen et al. (2010)]{Cuppen2010}
%H.M. Cuppen, S. Ioppolo, H. Linnartz
Cuppen, H.M., Ioppolo, S., \& Linnartz, H. 
%, Water formation at low temperatures by surface O$_2$ hydrogenation II: the reaction network, 
\ 2010, Phys.Chem.Chem.Phys., {12}, 12077 %(2010)
    
\bibitem[Crapsi et al. (2005)]{Crapsi2005}
%A. Crapsi, P. Caselli, C. M. Walmsley, P. C. Myers, M. Tafalla, C. W. Lee, and T. L. Bourke
Crapsi, A., Caselli, P., Walmsley, C.M., et al. 
%, Probing the Evolutionary Status of Starless Cores through N$_2$H$^+$ and N$_2$D$^+$ Observations,
\ 2005, \apj, {619}, 379 %(2005) Astrophys. J. , 

\bibitem[Crapsi et al. (2007)]{Crapsi2007}
%A. Crapsi, P. Caselli, M. C. Walmsley, and M. Tafalla
Crapsi, A., Caselli, P., Walmsley, M.C., \& Tafalla, M. 
%, Observing the gas temperature drop in the high-density nucleus of L 1544*, 
\ 2007, \aap, 470, 221 %(2007)   Astronomy \& Astrophysics,

\bibitem[Du et al. (2012)] {Du2012}
%F. Du,  B. Parise, and P. Bergman
Du, F., Parise, B., \& Bergman, P. 
%, Production of interstellar hydrogen peroxide (H$_2$O$_2$) on the surface of dust grains, 
\ 2012, \aap, {538}, A91 %Astronomy \& Astrophysics,

\bibitem[Dulieu et al. (2010)]{Dulieu2010}
%F. Dulieu, L. Amiaud, E. Congiu, J. H. Fillion, E. Matar, A. Momeni, V. Pirronello, J. L. Lemaire
Dulieu, F., Amiaud,  L., Congiu, E., et al. 
%, Experimental evidence for water formation on interstellar dust grains by hydrogen and oxygen atoms, 
\ 2010, \aap, {512}, A30 %(2010) Astron. Astrophys.,

\bibitem[Endres et al. (2016)]{Endres2016}
%C. P. Endres, S. Schlemmer, P. Schilke, J. Stutzki, and H. S. P. Müller
Endres, C.P., Schlemmer, S., Schilke,  P., et al. 
%, The Cologne Database for Molecular Spectroscopy, CDMS, in the Virtual Atomic and Molecular Data Centre, VAMDC, 
\ 2016, J. Mol. Spectrosc., {327}, 95 %(2016)

%
%
% check how to cite a book...
\bibitem[Engargiola \& Plambeck (1999)]{Engargiola1999}
Engargiola, G. , \& Plambeck, R.L.
\ 1999, The Physics and Chemistry of the Interstellar Medium, 
Eds.: V. Ossenkopf, J. Stutzki, and G. Winnewisser %(1999)

\bibitem[Fuchs et al. (2020)]{Fuchs2020}
%G.W. Fuchs, D. Witsch, D. Herberth, M. Kempkes, B. Stanclik, H. Linnartz, K. Menten, and T.F. Giesen
Fuchs, G.W., Witsch, D. , Herberth, D. , et al. 
%, Simulating young stellar object H$_2$CO and CH$_3$OH chemistry using an extended spherical physical-chemical model,
%%Testing a new physical-chemical model for Young Stellar Objects - Using the ice-borne molecules H$_2$CO and CH$-3$OH as case study,
Astron. Astrophys.,  (2020, submitted) 

\bibitem[Franklin et al. (2008)]{Franklin2008}
%Jonathan Franklin, Ronald L. Snell, Michael J. Kaufman, Gary J. Melnick, David A. Neufeld, David J. Hollenbach, and Edwin A. Bergin
Franklin, J.,  Snell, R.L., Kaufman,M.J.,  et al. 
%, SWAS Observations of Water in Molecular Outflows,
\ 2008, \apj, {674}, 1015 %(2008) Astrophysical J., 

\bibitem[Goldsmith \& Langer (1999)]{Goldsmith1999}
%Paul F. Goldsmith \& William D. Langer
Goldsmith, P.F., \&  Langer, W.D.
%, Population Diagram Analysis of Molecular Line Emission, 
\ 1999, \apj, {517}, 209 %(1999) Astrophysical J.,

\bibitem[Goldsmith et al. (2000)]{Goldsmith2000}
%P. F. Goldsmith, G. J. Melnick, E. A. Bergin, J. E. Howe, R. L. Snell, D. A. Neufeld, M. Harwit, M. L. N. Ashby, B. M. Patten, S. C. Kleiner, R. Plume, J. R. Stauffer, V. Tolls, Z. Wang, Y. F. Zhang, N. R. Erickson, D. G. Koch, R. Schieder, G. Winnewisser, and G. Chin
Goldsmith, P. F., Melnick, G. J., Bergin, E. A., et al. 
%, O$_2$ in Interstellar Molecular Clouds,
\ 2000, \apj, {539}, L123 %(2000) Astrophysical J.,

\bibitem[Goldsmith et al. (2011)]{Goldsmith2011}
%Paul F. Goldsmith, Rene Liseau, Tom A. Bell, John H. Black, Jo-Hsin Chen, David Hollenbach,
%Michael J. Kaufman, Di Li, Dariusz C. Lis, Gary Melnick {\em et al.}
Goldsmith, P.F., Liseau, R., Bell,T.A.,  et al. 
%, Herschel Measurements of Molecular Oxygen in Orion, 
\ 2011, \apj, {737}, 96 %(2011) Astrophys. J., 

\bibitem[G\"{u}sten et al. (2006)]{Gusten2006}
%R. G\"{u}sten, L. \AA. Nyman, P. Schilke, K. Menten, C. Cesarsky, and R. Booth
G\"{u}sten, R.,  Nyman, L.\AA., Schilke, P., et al. 
%, The Atacama Pathfinder EXperiment (APEX) – a new submillimeter facility for southern skies, 
\ 2006, \aap, {454}, L13 %(2006) Astronomy \& Astrophysics,  

\bibitem[Hollenbach et al. (2009)]{Hollenbach2009}
%David Hollenbach, Michael J. Kaufman, Edwin A. Bergin, and Gary J. Melnick
Hollenbach, D., Kaufman, M.J., Bergin, E.A., \& Melnick, G.J. 
%, Water, O$_2$, and ice in molecular clouds, 
\ 2009, \apj, {690}, 1497 % (2009) Astrophys.J., 

\bibitem[Hunt et al. (1965)]{Hunt1965}
%R.H. Hunt, R.A. Leacock, C.W. Peters, \& K.T. Hecht
Hunt, R.H. , Leacock, R.A. , Peters, C.W., \& Hecht, K.T. 
%, Internal-Rotation in Hydrogen Peroxide: The Far-Infrared Spectrum and the Determination of the Hindering Potential,
\ 1965, \jcp, {42}, 1931 %(1965)

\bibitem[Ivezic \& Elitzur (1997)]{Ivezic1997}
%Z. Ivezic \& M, Elitzur
Ivezic, Z., \& Elitzur, M. 
%, Self-similarity and scaling behaviour of infrared emission from radiatively heated dust — I. Theory,
\ 1997, \mnras,  {287}, 799 %(1997) Mon. Notices Royal Astron. Soc.,

\bibitem[Ioppolo et al. (2008)]{Ioppolo2008}
%S. Ioppolo, H. M. Cuppen, C. Romanzin, E. F. van Dishoeck and H. Linnartz
Ioppolo, S., Cuppen, H. M., Romanzin, C., et al.
%, Laboratory Evidence for Efficient Water Formation in Interstellar Ices,
\ 2008, \apj,  {686},  1474 %(2008) Astrophys. J.,

\bibitem[Ioppolo et al. (2010)] {Ioppolo2010}
%S. Ioppolo, H.M. Cuppen, C. Romanzin, E.F. van Dishoeck, H.  Linnartz
Ioppolo, S., Cuppen, H.M., Romanzin, C., et al. 
%, Water formation at low temperatures by surface $O_2$ hydrogenation I: characterization of ice penetration, 
\ 2010, Phys.Chem.Chem.Phys., {12}, 12077 %(2010)
%\url{http://dx.doi.org/10.1039/C0CP00251H},

\bibitem[Jørgensen et al. (2004a)]{Jorgensen2004a}
%J.K. Jørgensen, M.R. Hogerheijde, E.F. van Dishoeck, G.A. Blake  \& F.L. Schöier
Jørgensen, J.K., Hogerheijde, M.R., van Dishoeck, E.F., et al. 
%, The structure of the NGC 1333-IRAS2 protostellar system on 500 AU scales - An infalling envelope, a circumstellar disk, multiple outflows, and chemistry,
\ 2004, \aap, {413}, 993 % (2004) Astronomy \& Astrophysics, 

\bibitem[Jørgensen et al. (2004b)]{Jorgensen2004b}
%J. K. Jørgensen, M. R. Hogerheijde, G. A. Blake, E. F. van Dishoeck, L. G. Mundy, and F. L. Schöier
Jørgensen, J.K., Hogerheijde, M.R., Blake, G.A., et al. 
%, The impact of shocks on the chemistry of molecular clouds - High resolution images of chemical differentiation along the NGC 1333–IRAS\,2A outflow, 
\ 2004, \aap, {415}, 1021 % (2004) Astronomy \& Astrophysics, 

\bibitem[Kristensen et al. (2010)] {Kristensen2010} 
%L. E. Kristensen, E. F. van Dishoeck, T.A. van Kempen, H. M. Cuppen, C. Brinch, J. K. Jørgensen, and M. R. Hogerheijde
Kristensen, L.E., van Dishoeck, E.F., van Kempen, T.A., et al. 
%, Methanol maps of low-mass protostellar systems - I. The Serpens molecular core, 
\ 2010, \aap, {516}, A57 %(2010) Astronomy \& Astrophysics,

\bibitem[Kristensen et al. (2012)] {Kristensen2012} 
%L. E. Kristensen, E. F. van Dishoeck, E. A. Bergin, R. Visser, U. A. Yıldız, I. San Jose-Garcia, J. K. Jørgensen, G. J. Herczeg, D. Johnstone, S. F. Wampfler, A. O. Benz, S. Bruderer, S. Cabrit, P. Caselli, S. D. Doty, D. Harsono, F. Herpin, M. R. Hogerheijde, A. Karska, T. A. van Kempen, R. Liseau, B. Nisini, M. Tafalla, F. van der Tak and F. Wyrowski
Kristensen, L.E., van Dishoeck, E.F., Bergin, E.A., et al. 
%, Water in star-forming regions with Herschel (WISH). II. Evolution of 557 GHz 110-101 emission in low-mass protostars, 
\ 2012, \aap, {542}, A8 %(2012) Astronomy \& Astrophysics, 

\bibitem[Larsson et al. (2007)]{Larsson2007}
%B. Larsson, R. Liseau, L. Pagani, P. Bergman, P. Bernath, N. Biver, J. H. Black, R. S. Booth, V. Buat, J. Crovisier et al. 
Larsson, B., Liseau, R., Pagani, L., et al. 
% Molecular oxygen in the $\rho$ Ophiuchi cloud,
\ 2007, \aap, {466}, 999 % (2007) Astronomy \& Astrophysics,

\bibitem[Larsson et al. (2017)]{Larsson2017}
%B. Larsson and R. Liseau
Larsson, B. \& Liseau, R. 
%, Gas and dust in the star-forming region $\rho$ Oph A - II. The gas in the PDR and in the dense cores,
\ 2017, \aap, {608}, A133 %(2017) Astronomy \& Astrophysics, 

\bibitem[Lee et al. (2014)]{Lee2014}
%Jeong-Eun Lee, Jinhee Lee, Seokho Lee, Neal J. Evans II, and Joel D. Green
Lee, J.-E., Lee, J., Lee, S., et al. 
%, Herschel Key Program, ``Dust, Ice, and Gas in Time'' (DIGIT): The Origin of Molecular and Atomic Emission in Low-Mass Protostars in Taurus,
\ 2014, \apjs, {214}, 21 % (2014) Astrophys. J. Suppl. Ser.,

\bibitem[Lefloch et al. (2010)]{Lefloch2010}
%B. Lefloch, S. Cabrit, C. Codella et al.
Lefloch, B., Cabrit, S., Codella, C., et al.
%, The CHESS spectral survey of star forming regions: Peering into the protostellar shock L1157-B1*, II. Shock dynamics,
\ 2010, \aap, {518}, L113 %(2010) Astronomy \& Astrophysics, 

\bibitem[Lindberg et al. (2017)]{Lindberg2017}
%Johan E. Lindberg, Steven B. Charnley, Jes K. Jørgensen, Martin A. Cordiner, and Per Bjerkeli
Lindberg, J.E., Charnley,  S.B.,  Jørgensen, J.K., et al. 
%, Externally Heated Protostellar Cores in the Ophiuchus Star-Forming Region,  
\ 2017, \apj, {835}, 3  % (2017) Astrophys. J., 

\bibitem[Lindberg et al. (2015)]{Lindberg2015}
%J. E. Lindberg, J. K. Jørgensen, Y. Watanabe, S. E. Bisschop, N. Sakai, and S. Yamamoto
Lindberg, J. E., Jørgensen, J. K., Watanabe, Y., et al. 
%, Probing the effects of external irradiation on low-mass protostars through unbiased line surveys, 
\ 2015, \aap, {584}, A28 %(2015) Astronomy \& Astrophysics, 

\bibitem[Liseau et al. (2012)] {Liseau2012}
%R. Liseau, P. F. Goldsmith, B. Larsson, L. Pagani, P. Bergman, J. Le Bourlot, T. A. Bell, A. O. Benz, E. A. Bergin, P. Bjerkeli, J. H. Black, S. Bruderer, P. Caselli, E. Caux, J.-H. Chen, M. de Luca, P. Encrenaz, E. Falgarone, M. Gerin, J. R. Goicoechea, Å. Hjalmarson, D. J. Hollenbach, K. Justtanont, M. J. Kaufman, F. Le Petit, D. Li, D. C. Lis, G. J. Melnick, Z. Nagy, A. O. H. Olofsson, G. Olofsson, E. Roueff, Aa. Sandqvist, R. L. Snell, F. F. S. van der Tak, E. F. van Dishoeck, C. Vastel, S. Viti and U. A. Yıldız
Liseau, R., Goldsmith, P. F., Larsson, B., et al. 
%, Multi-line detection of O$_2$ toward $\rho$ Ophiuchi A,
\ 2012, \aap, {541}, A73 %(2012) Astronomy \& Astrophysics,

\bibitem[Liseau \& Larsson (2015)]{Liseau2015}
%R. Liseau and B. Larsson
Liseau, R., \& Larsson, B. 
%, Search for HOOH in Orion, 
\ 2015, \aap, {583}, A53 %(2015) Astronomy \& Astrophysics,

\bibitem[Luspay-Kuti et al. (2018)]{Luspay-Kuti2018}
%A. Luspay-Kuti, O. Mousis, J.I. Lunine, Y. Ellinger, F. Pauzat, U. Raut, A. Bouquet, K.E. Mandt, R. Maggiolo, T. Ronnet, B. Brugger, O. Ozgurel, S. A. Fuselier
Luspay-Kuti, A., Mousis, O.,  Lunine, J.I., et al. 
%, Origin of Molecular Oxygen in Comets: Current Knowledge and Perspectives, 
\ 2018, \ssr, {214}, 115 %(2018) Space Sci. Rev., 


\bibitem[McMullin et al. (2007)]{McMullin2007}
%McMullin, J. P., Waters, B., Schiebel, D., Young, W., \& Golap, K., 
McMullin, J. P., Waters, B., Schiebel, D., et al. 
\ 2007, Astronomical Data Analysis Software and Systems XVI, eds. R. A. Shaw, F. Hill, \& D. J. Bell (San Francisco, CA: ASP), ASP Conf. Ser., {376}, 127 %(2007) 

\bibitem[Melnick \& Bergin (2005)]{Melnick2005}
%G.J. Melnick, E.A. Bergin
Melnick, G.J., \& Bergin, E.A. 
%, The legacy of SWAS: Water and molecular oxygen in the interstellar medium,
\ 2005, Advances in Space Research, {36}, 1027 %,  (2005)

\bibitem[Miyauchi et al. (2008)]{Miyauchi2008}
%N. Miyauchi, H. Hidaka, T. Chigai, A. Nagaoka, N. Watanabe, A. Kouchi
Miyauchi, N., Hidaka, H., Chigai, T., et al. 
%, Formation of hydrogen peroxide and water from the reaction of cold hydrogen atoms with solid oxygen at 10 K,
\ 2008, Chem. Phys. Lett., {456}, 27 %(2008)

\bibitem[Mottram et al. (2014)]{Mottram2014}
%J. C. Mottram, L.E.Kristensen, E. F. van Dishoeck, S.Bruderer, I. San José-García, A. Karska, R. Visser, G. Santangelo, A.O.Benz, E.A.Bergin, P. Caselli, F.Herpin, M.R.Hogerheijde, D. Johnstone, T. A . van Kempen, R. Liseau, B. Nisini, M. Tafalla, F.F.S.vanderTak, and F. Wyrowski
Mottram, J. C., Kristensen, L.E., van Dishoeck, E. F., et al. 
%, Water in star-forming regions with Herschel (WISH) V. The physical conditions in low-mass protostellar outflows revealed by multi-transition water observations, 
\ 2014, \aap, {572}, A21 % (2014) Astronomy \& Astrophysics, 

\bibitem[Minissale et al. (2016)]{Minissale2016}
%M. Minissale, F. Dulieu, S. Cazaux, and S. Hocuk
Minissale, M., Dulieu, F., Cazaux, S., \& Hocuk, S. 
%, Dust as interstellar catalyst I. - Quantifying the chemical desorption process, 
\ 2016, \aap, {585}, A24 % (2016) Astronomy \& Astrophysics, 
%DOI: 10.1051/0004-6361/201525981

\bibitem[Möller et al. (2017)]{Moller2017}
%T. Möller, C. Endres, and P. Schilke
Möller, T., Endres, C., \& Schilke, P. 
%, eXtended CASA Line Analysis Software Suite (XCLASS), 
\ 2017, \aap, {598}, A7 %(2017) Astronomy \& Astrophysics,

\bibitem[Nisini et al. (2005)]{Nisini2005}
%B. Nisini, S.  Antoniucci, T.  Giannini, and D. Lorenzetti
Nisini, B., Antoniucci, S., Giannini, T., \&  Lorenzetti, D.
%, Probing the embedded YSOs of the R CrA region through VLT-ISAAC spectroscopy, 
\ 2005, \aap, {429}, 543 %(2005) Astronomy \& Astrophysics, 

\bibitem[Nutter et al. (2005)] {Nutter2005}
%D.J.~Nutter, D.~Ward-THompson, and P.~Andre
Nutter, D.J., Ward-Thompson, D.,  \& Andre, P.
%, The pre-stellar and protostellar population of R Coronae Australis,
\ 2005, \mnras, {357}, 975-982 %(2005) Mon.Not.R.Astron.Soc.

\bibitem[Oba et al. (2009)]{Oba2009}
%Y. Oba, N. Miyauchi, H. Hidaka, T. Chigai, N. Watanabe, and A. Kouchi
Oba, Y., Miyauchi, N., Hidaka, H., et al. 
%, Formation of Compact Amorphous H$_2$O Ice by Codeposition of Hydrogen Atoms with Oxygen Molecules on Grain Surfaces,
\ 2009, \apj, {701}, 464-470 %, (2009) 

\bibitem[Öberg et al. (2009)]{Oberg2009}
%Karin I. Öberg, Harold Linnartz, Ruud Visser, and Ewine F. van Dishoeck
Öberg, K.I., Linnartz, H., Visser, R., \& van Dishoeck,  E.F. 
%, Photodesorption of Ices. II. H$_2$O AND D$_2$O,
\ 2009, \apj, {693}, 1209 % (2009) Astrophys. J.,

\bibitem[Osorio et al. (2003)]{Osorio2003}
%M. Osorio,P. D’Alessio, J. Muzerolle, N. Calvet and L. Hartmann
Osorio, M., D’Alessio, P., Muzerolle, J., et al. 
%, A Comprehensive Study of the L1551-IRS\,5 Binary System,
%% A COMPREHENSIVE STUDY OF THE L1551-IRS 5 BINARY SYSTEM,
\ 2003, \apj, {586}, 1148 %, (2003) Astrophys. J.,

\bibitem[Parise et al. (2012)] {Parise2012}
%B. Parise, P. Bergman, F. Du
Parise, B., Bergman, P., \& Du, F. 
%, Detection of the hydroperoxyl radical HO$_2$ toward $\rho$ Ophiuchi A: Additional constraints on the water chemical network,  
\ 2012, \aap, {541}, L11 %(2012) Astronomy \& Astrophysics, 

\bibitem[Parise et al. (2014)] {Parise2014}
%B. Parise, P. Bergman and K. Menten
Parise, B., Bergman P., \& Menten, K. 
%, Characterizing the chemical pathways for water formation – a deep search for hydrogen peroxide, 
\ 2014, Faraday Discuss., {168}, 349 %(2014)

\bibitem[Peterson et al. (2011)]{Peterson2011}
%Dawn E. Peterson, Alessio Caratti o Garatti, Tyler L. Bourke, Jan Forbrich, Robert A. Gutermuth, Jes K. Jørgensen, Lori E. Allen, Brian M. Patten, Michael M. Dunham, Paul M. Harvey, Bruno Merin, Nicholas L. Chapman, Lucas A. Cieza, Tracy L. Huard, Claudia Knez, Brian Prager, and Neal J. Evans II
Peterson, D.E., Caratti o Garatti, A., Bourke, T.L.,  et al. 
%, The Spitzer Survey of Interstellar Clouds in the Gould Belt. III. A Multi-Wavelength View of Corona Australis,  
%%THE SPITZER SURVEY OF INTERSTELLAR CLOUDS IN THE GOULD BELT. III. A MULTI-WAVELENGTH VIEW OF CORONA AUSTRALIS,  
\ 2011, \apjs, {194}, 43 %(2011) Astrophys. J. Suppl. Ser., 

\bibitem[Petkie et al. (1995)]{Petkie1995}
%D. T. Petkie, T. M. Goyette, J. J. Holton, F. C. De Lucia and P. Helminger
Petkie, D. T., Goyette, T. M., Holton, J. J., et al. 
%, Millimeter/Submillimeter-Wave Spectrum of the First Excited Torsional State in HOOH,
\ 1995, J. Mol.  Spect., {171}, 145 %(1995).

\bibitem[Pickett et al. (1998)]{Pickett1998} 
%H. M. Pickett, R. L. Poynter, E. A. Cohen, M. L. Delitsky, J. C. Pearson, and H. S. P. M\''{u}ller
Pickett, H. M., Poynter, R. L., Cohen, E. A., et al. 
%, Submillimeter, Millimeter, and Microwave Spectral Line Catalog, 
\ 1998, J. Quant. Spectrosc. \& Rad. Transfer, {60}, 883-890 %(1998) 

\bibitem[Romanzin et al. (2011)]{Romanzin2011}
%C. Romanzin, S. Ioppolo, H.M. Cuppen, E.F. van Dishoeck, H. Linnartz
Romanzin, C., Ioppolo, S., Cuppen, H.M., et al. 
%, Water formation by surface O$_3$ hydrogenation,
\ 2011, \jcp, {134}, 084504 %(2011) J.Chem.Phys, 

\bibitem[Sandell et al. (1994)]{Sandell1994}
%G. Sandell, L.B.G. Knee, C. Aspin, I.E. Robson, A.P.G. Russell
Sandell, G., Knee,  L.B.G., Aspin, C. , et al. 
%, A molecular jet and bow shock in the low mass protostellar binary NGC 1333-IRAS\,2, 
\ 1994, \aap, {285}, L1 %(1994) Astronomy \& Astrophysics, 

\bibitem[Schmalzl et al. (2014)]{Schmalzl2014}
%M. Schmalzl, R. Visser, C. Walsh, T. Albertsson, E.F.vanDishoeck, L. E. Kristensen, and J. C. Mottram
Schmalzl, M., Visser, R., Walsh, C., et al. 
%, Water in low-mass star-forming regions with Herschel - The link between water gas and ice in protostellar envelopes,
\ 2014, \aap, {572}, A81 %(2014) Astronomy \& Astrophysics,

\bibitem[Schoeier et al. (2002)]{Schoeier2002}
%F.L. Schöier, J.K. Jørgensen, E.F. van Dishoeck, \& G.A. Blake
Schöier, F.L., Jørgensen, J.K. , van Dishoeck, E.F. , \& Blake, G.A. 
%, Does IRAS\,16293–2422 have a hot core? Chemical inventory and abundance changes in its protostellar environment,
\ 2002, \aap, {390}, 1001 %(2002) Astronomy \& Astrophysics, 

\bibitem[Shirley et al. (2000)]{Shirley2000}
%Yancy L. Shirley, Neal J. Evans II, Jonathan M. C. Rawlings, and Erik M. Gregersen
Shirley, Y.L., Evans II, N.J., Rawlings, J.M.C., \& Gregersen, E.M. 
%, Tracing the Mass during Low-Mass Star Formation. I. Submillimeter Continuum Observations,
\ 2000, \apjs, {131}, 249 %(2000) Astrophys. J. Suppl. Ser., 

\bibitem[Smith et al. (2011)]{Smith2011}
%R. G. Smith, S. B. Charnley, Y. J. Pendleton, C. M. Wright, M. M. Maldoni, and G. Robinson
Smith, R.G., Charnley, S.B., Pendleton, Y.J., et al. 
%, On the Formation of Interstellar Water Ice: Constraints from a Search for Hydrogen Peroxide Ice in Molecular Clouds,
\ 2011, \apj, {743}, 131 %(2011) Astrophys. J., 

\bibitem[Stahl \& Palla (2005)]{Stahl2005}
%S.W. Stahler \& F. Palla
Stahler, S.W., \& Palla, F. 
\ 2005, The Formation of Stars, Wiley-VCH %(2005)

\bibitem[Tafalla et al. (1998)]{Tafalla1998}
%M. TAFALLA, D. MARDONES, AND P. C. MYERS, P. CASELLI, R. BACHILLER, P. J. BENSON, 
Tafalla, M., Mardones, D., Myers, P. C., et al.
%, L1544: A Starless Dense Core with Extended Inward Motions, 
%L1544: A STARLESS DENSE CORE WITH EXTENDED INWARD MOTIONS, 
\ 1998, \apj, {504}, 900 % (1998) Astrophys. J., 

\bibitem[Taylor \& Storey (1984)]{Taylor1984}
%K.N.R. Taylor \& J.W.V. Storey
Taylor, K.N.R., \& Storey, J.W.V. 
%, The Coronet, an obscured cluster adjacent to R Corona Austrina, 
\ 1984, \mnras, {209}, 5 %(1984) Mon. Not. R. astr. Soc.,

\bibitem[Tielens (2000)]{Tielens2000}
Tielens, A.G.G.M. 
%, Far-Infrared Spectroscopy of Interstellar Dust, 
\ 2000, Proc. 'The dusty and Molecular Universe', Paris, France, ESA SP-{577}, 245 %(2000)

\bibitem[Van den Bussche et al. (1999)]{VandenBussche1999}
%B. Vandenbussche1, P. Ehrenfreund2, A.C.A. Boogert4, E.F. van Dishoeck2,3, W.A. Schutte2, P.A. Gerakines5, J. Chiar7,A.G.G.M. Tielens4, J. Keane4, D.C.B. Whittet5, M. Breitfellner6, and M. Burgdorf
Vandenbussche, B., Ehrenfreund, P., Boogert, A.C.A., et al.
\ 1999, \aap, {346}, L57 %(1999) Astron.Astrophys., 

\bibitem[van Dishoeck et al. (2011)]{vanDishoeck2011}
%E. F. van Dishoeck, L. E. Kristensen, A. O. Benz, E. A. Bergin, P. Caselli, J. Cernicharo, F. Herpin, M. R. Hogerheijde, D. Johnstone, R. Liseau, B. Nisini, R. Shipman, M. Tafalla, F. van der Tak, F. Wyrowski, Y. Aikawa, R. Bachiller, A. Baudry, M. Benedettini, P. Bjerkeli, G. A. Blake, S. Bontemps, J. Braine, C. Brinch, S. Bruderer, L. Chavarría, C. Codella, F. Daniel, Th. de Graauw, E. Deul, A. M. di Giorgio, C. Dominik, S. D. Doty, M. L. Dubernet, P. Encrenaz, H. Feuchtgruber, M. Fich, W. Frieswijk, A. Fuente, T. Giannini, J. R. Goicoechea, F. P. Helmich, G. J. Herczeg, T. Jacq, J. K. Jørgensen, A. Karska, M. J. Kaufman, E. Keto, B. Larsson, B. Lefloch, D. Lis, M. Marseille, C. McCoey, G. Melnick, D. Neufeld, M. Olberg, L. Pagani, O. Panić, B. Parise, J. C. Pearson, R. Plume, C. Risacher, D. Salter, J. Santiago-García, P. Saraceno, P. Stäuber, T. A. van Kempen, R. Visser, S. Viti, M. Walmsley, S. F. Wampfler, and U. A. Yıldız
van Dishoeck, E.F., Kristensen, L.E., Benz, A.O.,  et al. 
%, Water in Star-forming Regions with the Herschel Space Observatory (WISH). I. Overview of Key Program and First Results, 
\ 2011, \pasp, {123}, 138 %(2011) Publ. Astron. Soc. Pacif., 

\bibitem[van Dishoeck et al. (2013)]{vanDishoeck2013}
%E.F. van Dishoeck, E. Herbst, D.A. Neufeld
van Dishoeck, E.F., Herbst, E., \& Neufeld, D.A. 
%, Interstellar Water Chemistry: From Laboratory to Observations, 
\ 2013, Chem. Rev., {113}, 9043 %(2013) 
 
\bibitem[Velilla-Prieto et al. (2017)]{Velilla-Prieto2017} 
%L. Velilla Prieto, C. Sánchez Contreras, J. Cernicharo, M. Agúndez, G. Quintana-Lacaci, V. Bujarrabal, J. Alcolea, C. Balança, F. Herpin, K. M. Menten, and F. Wyrowski
Velilla Prieto, L., Sánchez Contreras, C., Cernicharo, J., et al. 
%, The millimeter IRAM-30m line survey toward IK Tauri, 
\ 2017, \aap, {597}, A25 %(2017) Astronomy \& Astrophysics,
%DOI: 10.1051/0004-6361/201628776
 
\bibitem[Villa et al. (2017)]{Villa2017}
%A. M. Villa, M. A. Trinidad, E. de la Fuente, and T. Rodriguez-Esnard
Villa, A.M., Trinidad, M.A., de la Fuente, E., \& Rodriguez-Esnard, T. 
%, Proper Motions of L1551-IRS\,5 Binary System Using 7 mm VLA Observations,
%%PROPER MOTIONS OF L1551-IRS 5 BINARY SYSTEM USING 7 MM VLA OBSERVATIONS, 
\ 2017, \rmxaa, {53}, 525 %(2017) Revista Mexicana de Astronomia y Astrofisica,

\bibitem[Ward-Thompson et al. (1996)]{Ward-Thompson1996}
%D. Ward-Thompson, H.D. Buckley, J.S. Greaves, W.S. Holland and P. Andre
Ward-Thompson,D., Buckley, H.D., Greaves, J.S., et al. 
%, Evidence for protostellar infall in NGC 1333-IRAS\,2, 
\ 1996, \mnras, {281}, L53 %(1996) Mon.Not.R.Astron.Soc., 

 
\bibitem[White et al. (2000)]{White2000}
%G.J. White, R. Liseau, A.B. Men'shchikov, K. Justtanont, B. Nisini, M. Benedettini, E. Caux, C. Ceccarelli, J.C. Correia, T. Giannini, M. Kaufman, D. Lorenzetti, S. Molinari, P. Saraceno, H.A. Smith, L. Spinoglio, and E. Tommasi
White, G.J.,  Liseau, R., Men'shchikov, A.B., et al. 
%, An infrared study of the L1551 star formation region, 
\ 2000, \aap, {364}, 741 %(2000) Astronomy \& Astrophysics, 
 

\bibitem[Wilkin et al. (2002)]{Wilkin2002}
%F.P. Wilkin, H.-R. Chen, L.M. Chernin, \& R.L. Plambeck
Wilkin, F.P., Chen, H.-R., Chernin, L.M., \& Plambeck, R.L. 
%, SiO Observations of the NGC 1333 IRAS\,2A Protostellar Jet, 
\ 2002, arXiv:astro-ph/0212247
 
\bibitem[Wirström et al. (2016)]{Wirstrom2016}
%Eva S. Wirström, Steven B. Charnley, Martin A. Cordiner, and Cecilia Ceccarelli
Wirström, E.S., Charnley, S.B., Cordiner, M.A., \& Ceccarelli, C. 
%, A search for O$_2$ in co-depleted molecular cloud cores with HERSCHEL, 
\ 2016, \apj , {830}, 102 %(2016) Astrophys. J.

\bibitem[Yildiz et al. (2013)]{Yildiz2013}
%Umut A. Yıldız, Kinsuk Acharyya, Paul F. Goldsmith, Ewine F. van Dishoeck, Gary Melnick, Ronald Snell, René Liseau, Jo-Hsin Chen, Laurent Pagani, Edwin Bergin, Paola Caselli, Eric Herbst, Lars E. Kristensen, Ruud Visser, Dariusz C. Lis, and Maryvonne Gerin
Yıldız, U.A., Acharyya, K., Goldsmith, P.F., et al. 
%, Deep observations of O2 toward a low-mass protostar with Herschel-HIFI,
\ 2013, \aap, {558}, A58 %(2013) Astron.Astrophys., 



%-------------------------------------------------------------------

\end{thebibliography}
\end{document}